# Superhydrophobic sand mulches increase agricultural productivity in arid regions

Adair Gallo Jr.[1], Kennedy Odokonyero[1], Magdi A. A. Mousa[2,3], Joel Reihmer[1], Samir Al-Mashharawi[1], Ramona Marasco[4], Edelberto Manalastas[1], Mitchell J. L. Morton[4], Daniele Daffonchio[4], Matthew F. McCabe[1], Mark Tester[4], Himanshu Mishra[1]*

[1]King Abdullah University of Science and Technology, Water Desalination and Reuse Center, Division of Biological and Environmental Sciences and Engineering, Thuwal 23955-6900, Saudi Arabia

[2]Department of Arid Land Agriculture, Faculty of Meteorology, Environment and Arid Land Agriculture, King Abdulaziz University, 80208 Jeddah, Saudi Arabia

[3]Department of Vegetables, Faculty of Agriculture, Assiut University, 71526 Assiut, Egypt

[4]King Abdullah University of Science and Technology, Division of Biological and Environmental Sciences and Engineering, Thuwal 23955-6900, Saudi Arabia

*Corresponding author: Himanshu.Mishra@kaust.edu.sa

**Abstract**

Excessive evaporative loss of water from the topsoil in arid-land agriculture is compensated via irrigation, which exploits massive freshwater resources. The cumulative effects of decades of unsustainable freshwater consumption in many arid regions are now threatening food–water security. While plastic mulches can reduce evaporation from the topsoil, their cost and non-biodegradability limit their utility. In response, we report on superhydrophobic sand (SHS), a bio-inspired enhancement of common sand with a nanoscale wax coating. When SHS was applied as a 5 mm-thick mulch over the soil, evaporation dramatically reduced and crop yields increased. Multi-year field trials of SHS application with tomato (*Solanum lycopersicum*), barley (*Hordeum vulgare*), and wheat (*Triticum aestivum*) under normal irrigation enhanced yields by 17%–73%. Under brackish water irrigation (5500 ppm NaCl), SHS mulching produced 53%–208% higher fruit yield and grain gains for tomato and barley. Thus, SHS could benefit agriculture and city-greening in arid regions.

**Keywords:** food–water security; superhydrophobic sand; bio-inspiration; mulching; sustainable agrotechnology

The importance of irrigation toward humanity's ability to produce food cannot be overstated. For example, while only 20% of cultivated land is irrigated, this fraction contributes 33%–40% of the total world food production[1, 2]. Unfortunately, this outsized contribution to food production comes at a price, consuming over 70% of global freshwater withdrawals annually[2, 3, 4]. Regions with arid and semi-arid climates, such as the Middle East, northern Africa, the northwest Indian subcontinent, and western Australia, rely on their limited freshwater resources to grow food for sustenance and trade[3]. The plants growing in these regions depend on soil moisture content for nutrient uptake, optimal temperature regulation, and salt stress reduction[5]. However, due to intense solar radiation and direct exposure to dry air and winds, a significant fraction of the water supplied to soils is lost to evaporation (Fig. 1A)[6]. Therefore, to ensure sufficient water availability to support plant growth, excessive volumes of ground and river waters are routinely withdrawn; this has critically depleted water supplies in many parts of the world[7], resulting in food and water security concerns to become issues of international importance[8, 9, 10, 11, 12, 13].

Technologies for enhancing the water-use efficiency of irrigated agriculture, i.e., growing more food/biomass with less water, are warranted for a sustainable future. Subsurface impermeable layers are sometimes employed to limit water loss due to percolation; however, their installation is expensive and labor-intensive[14, 15]. As an alternative, the soil surface can be covered with plastic mulches to reduce evaporation, enhance crop yield, and reduce the incidence of pests and weeds as well as nutrient leaching[5, 16]. However, this approach is infrastructure-intensive and plastic landfilling is unsustainable[16]; harmful effects of residual plastic films in soils on crop yields have also been reported[17]. Biodegradable plastics are being vigorously pursued, but their success has been limited to date due to their high cost, slow or incomplete biodegradability, and time-varying wetting properties after deployment[16, 18, 19]. Recently, engineered nanomaterials (ENMs) have been demonstrated to enhance the water-use efficiency of soils[20]. For example, a pot-scale study revealed that an ENM comprising electron beam-dispersed attapulgite and sodium polyacrylate polyacrylamide reduced water and nutrient loss and boosted plant growth[21]. Nevertheless, to avoid unintended environmental consequences caused by the use of ENMs, multiscale transdisciplinary efforts have been undertaken to pinpoint their effects on soils and the microbiome[20, 22].

Based on the aforementioned concerns, in the present study, we developed an innovative approach to control evaporation from soils by combining (i) common sand, a material readily available in soils of arid regions, and (ii) paraffin wax, a low-cost and biodegradable hydrophobic material that is available at an industrial scale[23, 24, 25]. Our results demonstrate that superhydrophobic sand (SHS) mulching reduces water evaporation from moist soils under authentic arid-land conditions (Fig. 1B) and that the enhanced soil moisture promotes plant health and yield.

**Results**

**Materials synthesis**

We drew inspiration from several super-water-repellent plants and animals that exploit waxy cocktails and micro- or nano-textures to achieve extreme water-repellency, also known as superhydrophobicity, to perform critical functions, including fog harvesting in arid conditions,[26] skating on and launching from the water surface to avoid predators,[27, 28] directing the movement of condensed water,[29, 30] and respiring under water[31]. Analogously, SHS comprises common sand grains coated with a nanoscale layer of paraffin wax (Figs. 1B and 2A). SHS was produced by dissolving common paraffin wax in hexane, mixing the solution with common sand, and evaporating hexane from the mixture at ~100 mbar and 55°C (Fig. S1). Then, hexane was simultaneously condensed and collected in a separate container for reuse during the process. The sand comprised silica particles 100–700 μm in diameter (Fig. 2B; particle-size distribution of a common sandy soil for comparison). Paraffin wax was a mixture of hydrocarbons with tails of 27–37 carbons (Fig. 2C, Fig. S2, Table S1). This process led to the formation of a 20-nm thick wax coating onto the sand grains, which we estimated from the mass, volume, and density of the wax, BET surface area of the sand (0.11 $m^2/g$), and wax-to-sand mass ratio of 1:600. In addition to hexane, pentane octane, cyclohexane, diethyl ether, dichloromethane, methyl-*t*-butyl ether, petroleum ether (ligroin), chloroform, tetrahydrofuran, and triethyl amine were tested as solvents; however, no significant differences were observed with respect to their water-repellent properties. Further, a variety of natural waxes were also tested, including palm wax, soy wax, and beeswax; the resultant sand exhibited superhydrophobicity. Given the scale of agricultural operations, we chose to use paraffin wax.

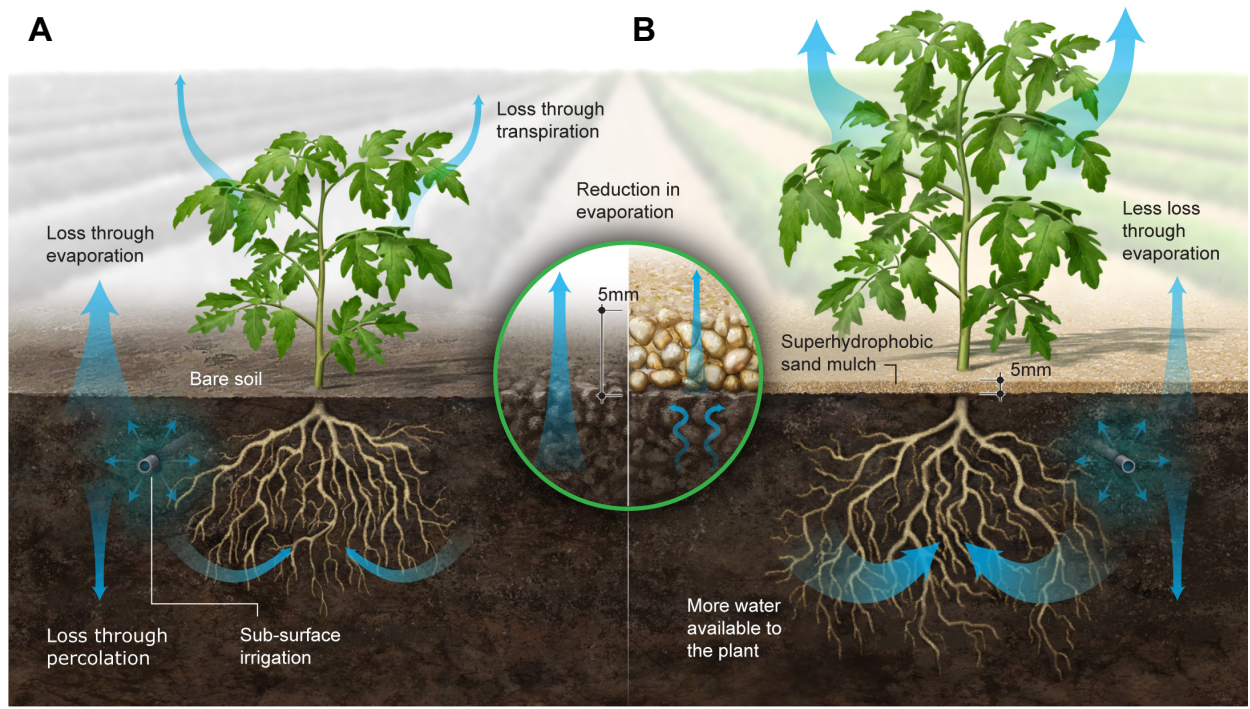

**Fig. 1 Concept of superhydrophobic sand (SHS) mulches to reduce water evaporation from soils in arid regions.** Water movements for subsurface irrigated (**A**) bare soil and (**B**) soil mulched with SHS. SHS prevents the capillary rise of water, thereby creating a dry diffusion barrier that allows water vapor to diffuse at a rate significantly lower than bare or unmulched soil.

**Water repellency**

The water repellency of sand grains dramatically increased after the above-described surface enhancement (Fig. 2A). The grain-level apparent contact angles, $\theta_r$, of water microdroplets formed by condensing water vapor on individual sand grains increased from $\theta_r \approx 30°$ (for ordinary sand) to $\theta_r \approx 105°$ for SHS (Fig. 2D-E). The advancing ($\theta_A$) and receding ($\theta_R$) contact angles of ≈10 µL of water droplets advanced and retracted at 0.2 µL/s on ~10 mm-thick SHS layers were $\theta_A \approx 160°$ and $\theta_R \approx 150°$, respectively, which are characteristics of superhydrophobicity (Methods). Water droplets of ~30 µL impacted a 5 mm-thick layer of SHS from a height of ~2 cm and bounced off and re-landed, forming liquid marbles[32] (Fig. 2F). The factors and mechanisms underlying the superhydrophobicity of SHS are presented in the Discussion section. Next, we characterized the breakthrough pressure of water on SHS, defined as the pressure at which water penetrates into the microtexture[33]. While water spontaneously imbibes into common sand, due to capillarity[34], a 5 mm-thick SHS layer could prevent imbibition of the water column up to a height of $h \leq 12$ cm, thereby presenting a breakthrough pressure of $P_h = \rho g h \approx 1.2$ kPa (as explained in the Discussion section). Based on these results, we envisioned that when placed on moist topsoil, SHS would create a capillary and diffusion barrier for soil moisture as well as insulate it from direct exposure to solar radiation, wind, and dry air (Fig. 1B).

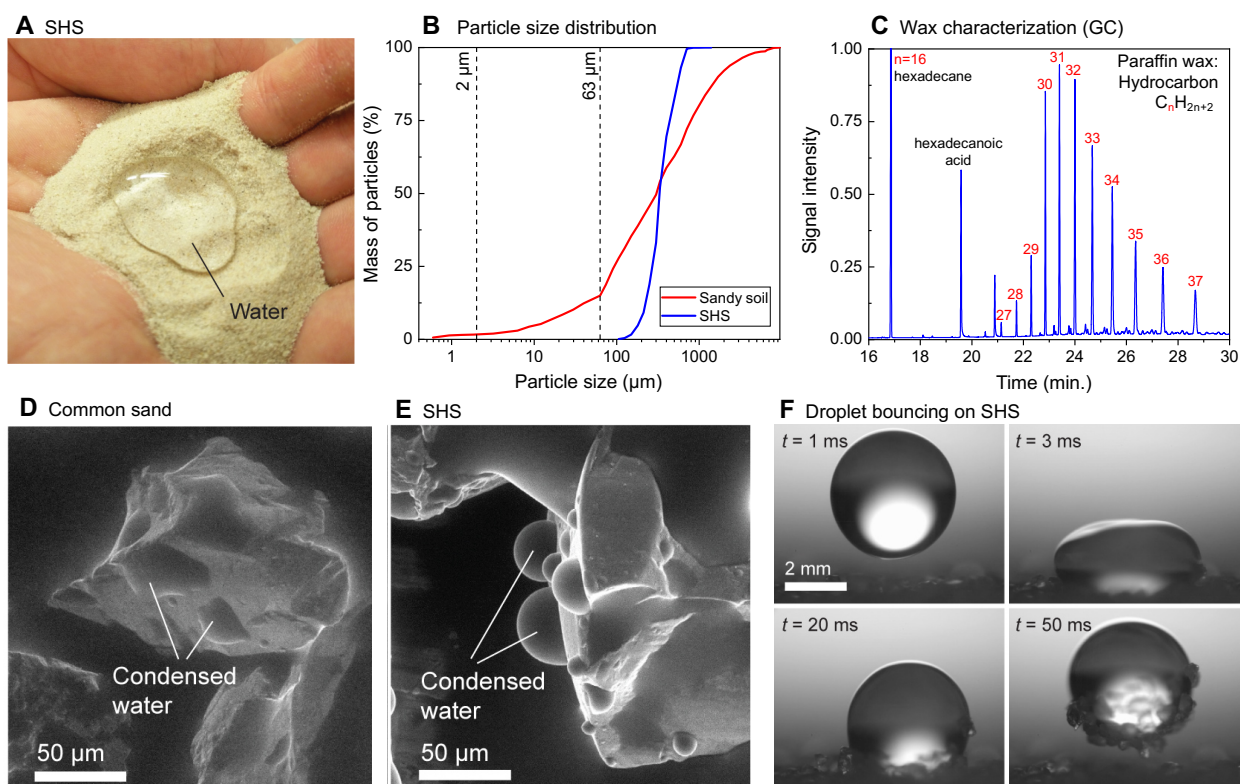

**Fig. 2 Characterization of superhydrophobic sand (SHS).** (**A**) (**C**) A photograph of SHS with water on top, demonstrating its superhydrophobicity (**B**) Particle-size distribution of SHS and sandy soil (loamy sand as per USDA soil texture characterization) collected at a local agriculture research facility. The fractions above and below 63 µm were determined using sieving and the hydrometer method[35], respectively. (**C**) Gas chromatography of paraffin wax (~0.1 M in cyclohexane) pinpointing the chain lengths of the constituent alkanes (n) in red. The relative compositions of the alkanes are presented in Table S1. Note: hexadecane and hexadecanoic acid were used as internal standards. Representative environmental scanning electron micrographs of water droplets condensed on (**D**) common sand grains and (**E**) SHS grains. The apparent contact angles of water droplets are significantly higher in (**E**) than in (**D**). (**F**) High-speed images of a 30 µL water droplet dropped onto a 5 mm-thick SHS layer from a height of 2 cm (Movie S1).

**Effects of SHS mulching on evaporation flux and soil moisture content**

We evaluated the effects of SHS mulching on evaporation rates from the topsoil and actual moisture content and compared them with the effects of mulching with common sand and unmulched (control) sand. For this, bottom-closed pots were filled with 75 g potting soil and saturated with 130 g water. Some of these pots were then covered with either SHS or common sand, both with layer thicknesses ranging from 2.5–20.0 mm; the remaining unmulched pots were used as controls. The pots were exposed to day/night cycles for 26 days and the loss of soil moisture via evaporation was recorded (Fig. 3A-B). During this period, temperatures ranged from 23°C to 34°C and relative humidity ranged from 40% to 90% (Fig. 3C).

Compared to mulching with untreated sand and unmulched controls, SHS mulching dramatically decreased water loss via evaporation (Fig. 3D). During the first few days, while the soil moisture was still high, the evaporation flux of the pots mulched with untreated sand was high, similar to that of the controls (Fig. 3F). In contrast, the evaporation flux of SHS mulched pots was lower and mostly independent of soil moisture content (Fig. 3E). Remarkably, a 5 mm-thick SHS mulch layer reduced the evaporation flux from approximately 78% to 56% during the first 4 days of the experiment (Fig. 3F). Evaporation flux, $J$, was inversely proportional to SHS mulch thickness, $L_{SHS}$, such that $J \propto (L_{SHS})^{-1}$; as mulch thickness doubled, evaporative flux decreased by approximately 50% (Fig. 3G).

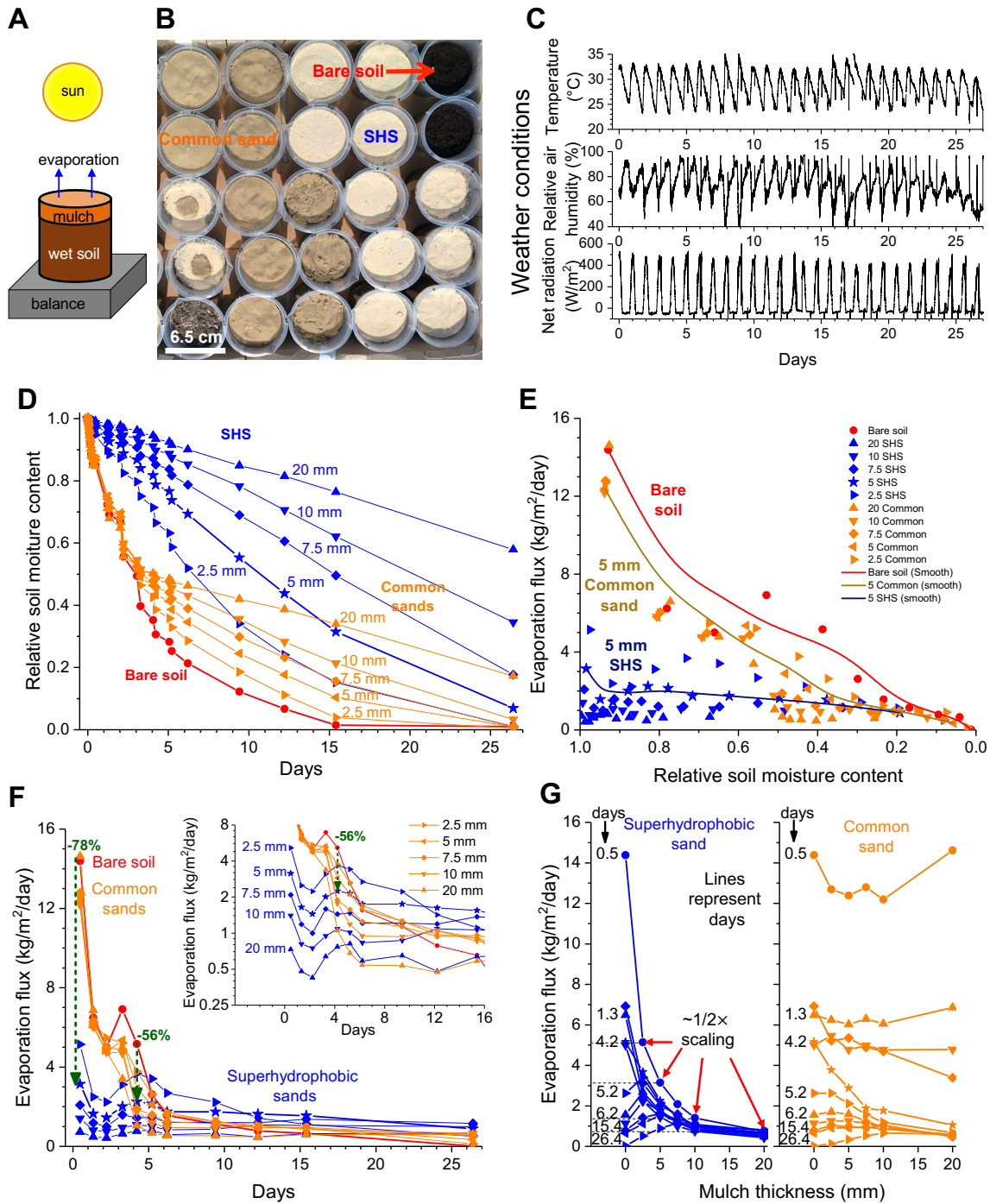

**Fig. 3 Effects of superhydrophobic sand (SHS) mulching on water loss via evaporation from soil.** We gravimetrically measured water loss from (bottom-closed) pots containing initially wet soil under various thicknesses of common sand and SHS mulches (0–20 mm, indicated by the number preceding the unit "mm") exposed to environmental conditions. SHS mulch reduced evaporation from the topsoil in proportion to its thickness. (**A**) Schematic of the experimental setup showing evaporation loss from a single pot. (**B**) Photograph of the pots on the first day reveal mulches with dark-colored common sand as they absorbed water from the soil underneath, whereas SHS stayed dry. (**C**) Weather conditions during the period of the experiment. (**D**)

Relative water loss from soil under different mulching conditions. (**E**) Evaporation flux as a function of relative soil moisture content. (**F**) Evaporation flux as a function of time. (**G**) Evaporation flux as a function of mulch thickness (Note: each curve represents the evaporative flux on a specific day as a function of mulch thickness; zero on the abscissa represents unmulched soil). The standard error was below 7% ($\sigma_{\bar{x}} = \sigma/\sqrt{N}$, where $\sigma$ is the standard deviation of the duplicates, $N = 2$).

Next, we quantified the effects of 5 mm-thick SHS mulches on soil moisture content at a farm with loamy sand soil at a field station located in Hada Al-Sham, Saudi Arabia (21.79° N, 39.72° E) and compared it with those of unmulched soil (control). Unlike the pot-scale experiment, this setup facilitated percolation. We installed soil moisture and temperature sensors that were specifically calibrated for this soil in bottom-perforated buckets that prevented the lateral flow of water while maintaining vertical flux. The buckets were then buried to soil level to minimize unrealistic soil temperature profiles during day/night cycles (Fig. 4A-C). We studied two irrigation scenarios: (i) single irrigation starting from supersaturated soil with no further water application (Fig. 4D) and (ii) daily irrigation using a subsurface drip system, also started from supersaturated soil (Fig. 4E, B). As both systems started from supersaturation, there were minimal differences in the beginning due to water loss via percolation in either case. However, as percolation abated, the effects of mulching on the soil moisture content became apparent: the moisture content in the top 5 cm of the unmulched soil was lost after approximately 14 days; in contrast, while it took approximately 23 days for mulched soil to dry out (Fig. 4D). Next, in the daily irrigation setup, soil moisture achieved steady states in approximately a week, such that the soil moisture of mulched soil was approximately 25%–45% higher than that of unmulched soil (Fig. 4E). Interestingly, the evaporation rates from unmulched supersaturated soil were quite high during the first few days, resulting in evaporative cooling, as evidenced by temperature data. These trends, however, reversed after a few days because evaporative cooling diminished at a steady state, such that the temperature of mulched soil was 1°C–3°C cooler than that of unmulched soil (Fig. 4D-E; bottom panels).

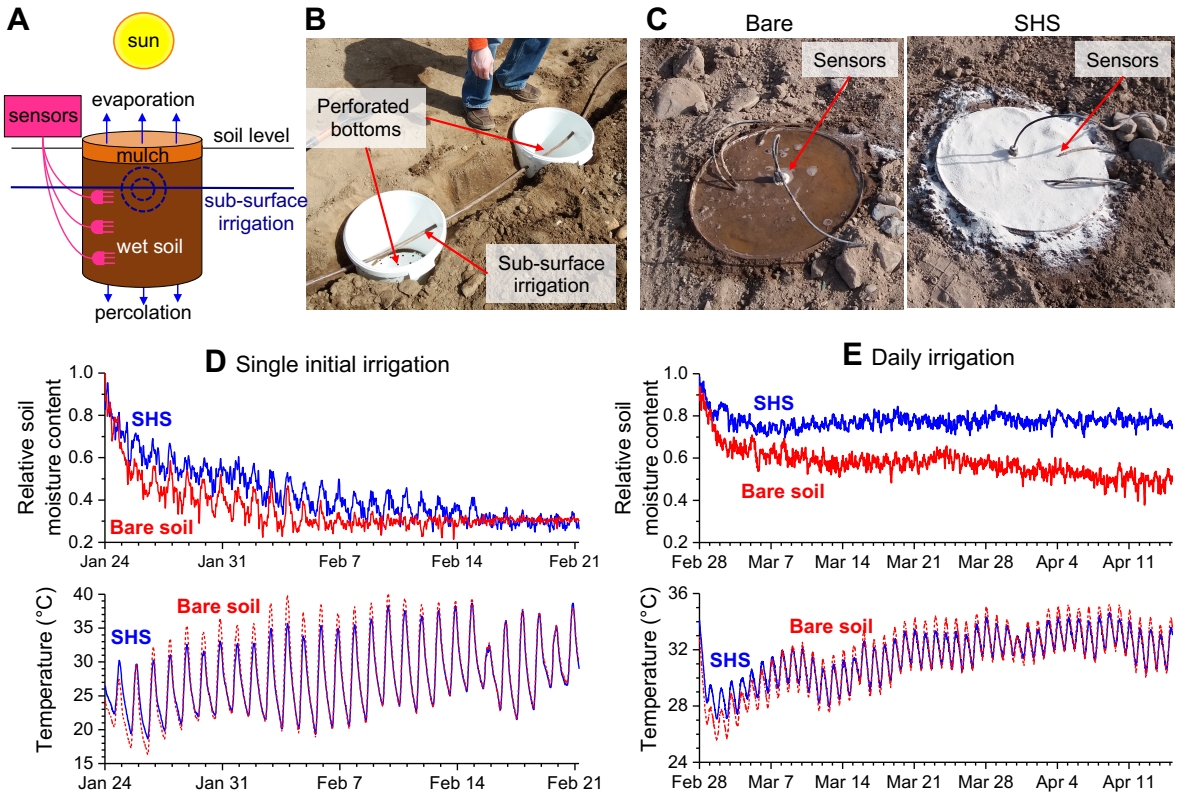

**Fig. 4 Effects of superhydrophobic sand (SHS) mulching on soil moisture content.** We used sensors to measure the effects of a 5 mm-thick SHS mulch layer on soil moisture content and soil temperature exposed to field conditions. (**A**) Schematic of the experimental setup showing evaporation loss and soil moisture/temperature sensors in a bottom-perforated bucket buried at soil level for a more realistic soil temperature profile. One irrigation line with a single dripping point passed through the buckets at a depth of approximately 10 cm. (**B-C**) Photographs of the experimental setup during installation and just after SHS application in one of the buckets. (**D-E**) Average soil moisture and temperature for two sets of experiments: (**D**) a single initial irrigation event up to saturation and another with (**E**) one initial irrigation to saturation followed by daily irrigation. In (**D**), the data were collected at a depth of 5 cm. In (**E**), the data represent the average of soil moisture and temperature from the following depths: 15, 20, and 25 cm. Soil moisture content significantly increased after SHS mulching, particularly in the case of daily irrigation. Mulching also reduced the highest daily temperatures, except during the first few days, when evaporative cooling was very high for unmulched soil.

**Effects of SHS mulching on crop yields**

Next, we investigated the effects of SHS mulching on agricultural productivity under true arid conditions at our field station located at Hada Al-Sham, Saudi Arabia (21.79° N, 39.72° E). Here, we conducted multiyear field trials using a high-value crop (tomato, *Solanum lycopersicum*

– variety A, cv. Bushra; variety B, cv. Nunhem's Tristar F1) and two large-scale grass crops (barley, *Hordeum vulgare*, cv. Morex and wheat, *Triticum aestivum*, cv. Balady). We compared the beneficial effects of SHS mulches of varying thicknesses with those of 120 μm-thick polyethylene sheets (hereafter referred to as plastic mulches) and unmulched soil (control). Two types of irrigation scenarios were investigated: (i) normal freshwater irrigation (<900 ppm NaCl) applied twice a day through subsurface drip irrigation and (ii) brackish irrigation (5000 ppm NaCl) applied twice a day (Fig. 5). Due to mild weather conditions, the crops were sown during the November–December period (winter 2017 and 2018) and were subsequently harvested during March–April of the following years (Figs. S4–S10).

Compared with unmulched controls, SHS mulching led to significant improvements in tomato yields under both normal and brackish water irrigation (Fig. 5A). The most significant result was obtained for the 5 mm-thick SHS mulch under normal irrigation for tomato, variety B (2018 season), which achieved a 72% improvement in yield relative to that obtained for unmulched controls. The yield enhancements for variety A with 5 mm and 10 mm-thick SHS mulches under normal irrigation were 27% and 40%, respectively, compared with those for unmulched controls. In fact, the performance of SHS mulching was on par with plastic mulches for tomato variety A, where black and clear plastic mulches enhanced crop yields by 28% and 43%, respectively. In our experiments with brackish water irrigation, the 5 mm-thick SHS mulch enhanced the yield of tomato variety B by 53% compared with unmulched controls (Fig. 5A). All of these results were statistically significant at $p < 0.05$ (Kruskal–Wallis H test).

We also observed statistically significant enhancements in the yield of barley mulched with the 5 mm-thick SHS mulch under normal (73%) and brackish water (208%) irrigation compared with unmulched controls (Fig. 5B). For wheat, SHS mulching led to a 17% enhancement in grain yield under normal fresh water irrigation. Under brackish water irrigation, only the dry biomass of barley plant that underwent SHS mulching was significantly higher (+44%) than that of barley plant grown in unmulched controls (Fig. S13). Mulching-enhanced soil moisture likely played a key role in transporting salts away from the root region via percolation/capillarity, as evidenced by the 39%–60% lower sodium concentration in the topsoil ($p < 0.05$, Table S2). Detailed insights into this mass transport necessitate further research, which falls beyond the scope of this study.

In 2020, we established an experiment with 450 tomato plants. However, this time, instead of assigning separate plots for specific treatments, we intercalated SHS and bare treatments in the same plots. These tomato plants produced a modest 10% increase in fruit yield compared with unmulched controls (Fig. S14). We consider that the experimental (intercalated) configuration led to the inadvertent sharing of the enhanced soil moisture beneath the SHS mulch with the neighboring unmulched regions/plants due to capillarity (note: the plants were separated by a distance of approximately 20–40 cm). Nevertheless, the results demonstrated statistical significance ($p = 0.02$) due to the greater number of replicates for this crop cycle.

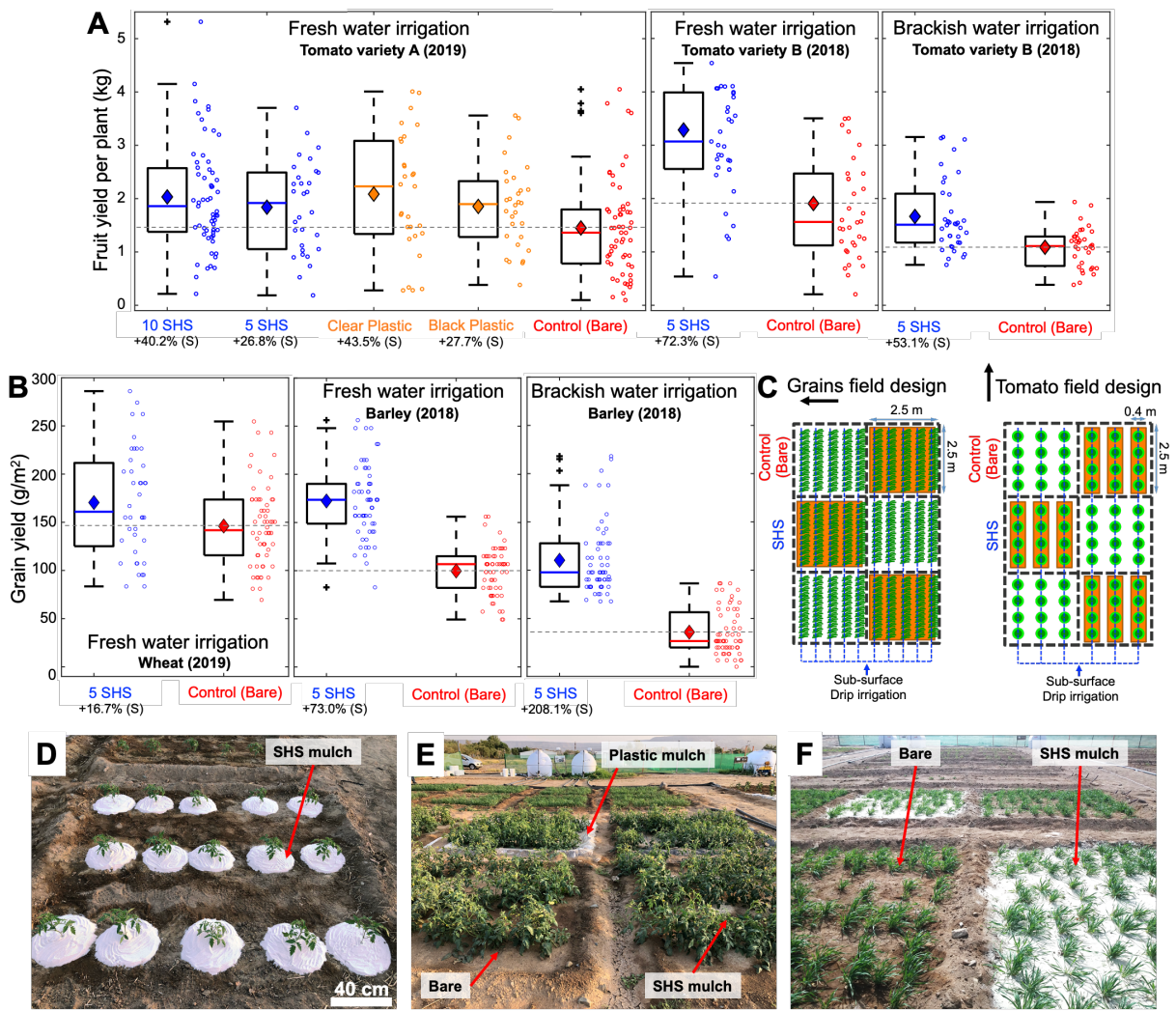

**Fig. 5 Effects of superhydrophobic sand (SHS) mulching on the performance of tomatoes, barley, and wheat in field trials.** Results for plants grown under 5–10 mm superhydrophobic sand (5–10 SHS) mulches (blue) versus bare soil (red) and plastic mulches (orange) on tomato

(**A**) fresh fruit yield and (**B**) barley and wheat grain yield. (**C**) Schematics of the field designs for grains with the SHS mulch completely covering the plot and tomato with the SHS mulch as 40-cm strips or circles around the plants. Photographs showing (**D–E**) tomato and (**F**) barley fields with arrows detailing the bare, SHS, and plastic mulch plots. SHS mulching significantly increased the yields of tomatoes, barley, and wheat. Dots in the boxplots represent the measurements for individual plants from replicate plots. Each dot represents the total mass of fruits produced per plant for tomatoes. The mass of grains per area is presented for barley and wheat. Each treatment was compared with the control (bare) case using the Kruskal–Wallis H test, where (S) represents statistical significance ($p < 0.05$) and percent change is the relative difference of the means. The boxes contain the middle 50% of the data points, with the horizontal line indicating the median and the diamond inside the box indicating the mean.

Lastly, out of curiosity, we investigated the effects of SHS mulching with reduced irrigation (50% of normal irrigation, once a day). Under this condition, the results were not promising, presumably due to the acute water stress (Figs. S11-12). This aspect needs to be investigated further.

**Effects of SHS mulching on soil and the root microbiome**

Unlike plastic mulches, which must be disposed of via landfilling[16], SHS mulches are environmentally benign. After crop harvesting in our field trials, we left SHS mulches in the field, where they lost their hydrophobicity in 7–9 months presumably due to the breakdown of wax by solar irradiation. Prior to the next crop cycle, we plowed the field and integrated the depleted SHS into the local sandy soil (for soil characterization, please refer to Supplementary Information, Section III, Fig. S15). To detect the impact of SHS mulching on the soil microbiome, we characterized the structure and composition of edaphic and plant root microbial communities of tomato and barley plants (2018 season) by collecting and analyzing root, rhizosphere, and bulk soil samples for unique sequence variants (SVs) of the bacterial 16S rRNA gene. We identified a total of 6,912 bacterial 16S rRNA gene unique SVs associated with root system compartments (root tissues and rhizosphere) and bulk soil samples of barley (6,338 SVs) and tomato (6,350 SVs; Table S5). Of the tested experimental factors (root system compartments, irrigation type, and the SHS mulch overlay), variation (beta-diversity) in bacterial community was mainly affected by niche-compartmentalization of the root system (barley: $R^2 = 0.26$, $F_{2,98} = 18.04$, $p = 0.001$ and tomato: $R^2 = 0.22$, $F_{2,99} = 14.99$, $p = 0.001$; Fig. 6A, D), followed by irrigation type, i.e., fresh versus brackish water, (barley: $R^2 = 0.04$, $F_{1,99} = 5.29$, $p = 0.001$ and tomato: $R^2 = 0.04$, $F_{1,100} = 5.76$, $p = 0.001$; Fig. 6B, E). The application of the SHS mulch overlay did not affect the composition of bacterial communities (barley: $R^2 = 0.011$, $F_{1,99} = 1.54$, $p = 0.06$ and tomato: $R^2 = 0.009$, $F_{1,100} = 1.27$, $p = 0.15$; Fig. 6C, F). Multiple comparison tests showed that in both crops, irrigation type induced a significant change in the composition of the bacterial communities associated with different plant root system compartments; in contrast, SHS mulch overlay had no effect on the composition of the bacterial communities (Table S6). This finding was supported by the number of SVs shared between the treatment with SHS mulches and normal soil, 4,954 (78%) for barley and 5,024 (79%) for tomato, accounting for 98% of the relative abundance in both crops, (Fig. 6G-H). *Gammaproteobacteria*, *Alphaproteobacteria* and *Bacteroidetes*, followed by *Firmicutes*, *Gemmatimonadetes* and *Actinobacteria* were dominant in the bacterial communities (Fig. S16).

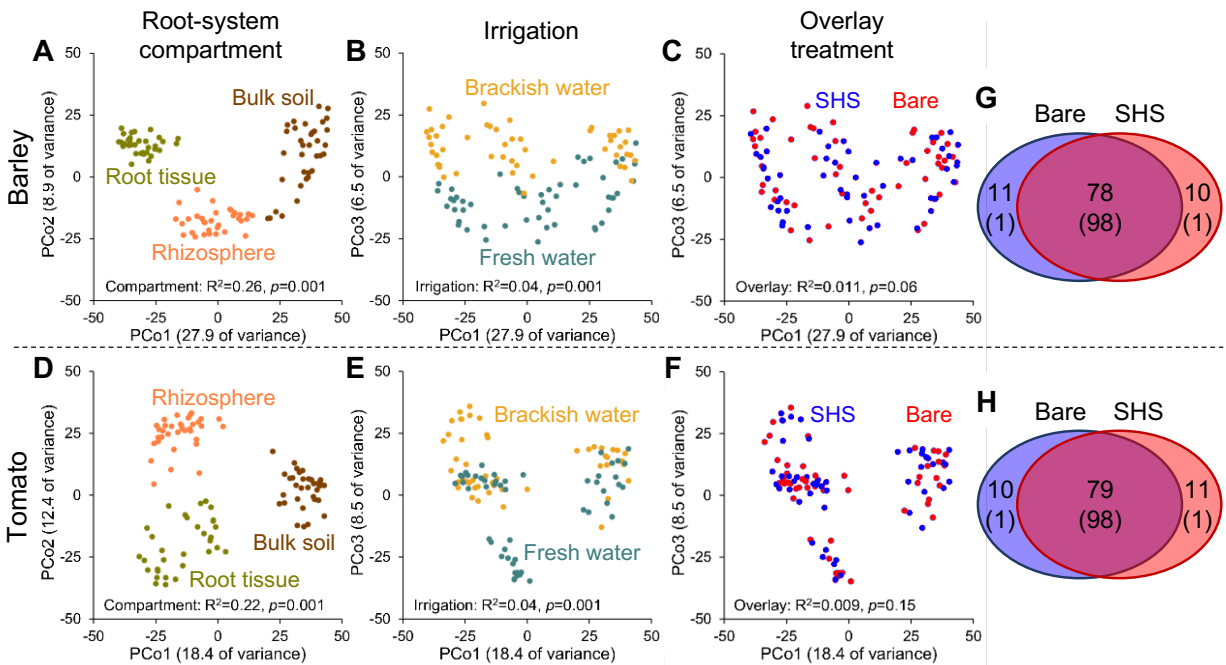

**Fig. 6 Unconstrained analysis of principal coordinates conducted on the root-system-associated bacterial communities.** We used the tested experimental factors (compartmentalization, irrigation type, and overlay treatment) to evaluate the variations in the overall composition of the bacterial communities in (**A-C**) barley and (**D-F**) tomato. Venn diagram shows the percentages of bacterial sequence variants (SV) shared among overlay treatments (bare soil and SHS mulch) in the root system (root and rhizosphere) and bulk soil of (**G**) barley and (**H**) tomato. SVs percentages are indicated by large areas, whereas the relative abundances of SVs are reported in parenthesis.

**Discussion**

The results of the present study underscore the potential of SHS mulching for growing more food and vegetation with limited freshwater resources in arid regions. In this section, we explain the mechanisms and factors underlying the water-repellency of SHS and its ability to reduce water loss via evaporation from the topsoil, followed by some arguments for the scalability of this approach.

First, we explain why micron-scale water droplets on individual SHS grains exhibit apparent contact angles of $\theta_o \approx 105°$ (Fig. 2E), whereas millimeter-scale water droplets placed on SHS mulches, e.g., a 5 mm-thick layer, exhibit much higher values, i.e., $\theta_A \approx 160°$ (Fig. 2A). This dramatic enhancement in water repellence arises from the entrapment of air between the SHS grains as they come into contact with water—a hallmark of superhydrophobicity[26, 27, 28, 29, 30, 31, 36]. In fact, the apparent macroscale angles can be related to grain-level (actual) angles via the Cassie–Baxter model, $\cos\theta_A = \phi_{LS} \times \cos\theta_o - \phi_{LV}$, where $\phi_{LS}$ and $\phi_{LV}$ are the area fractions of the real liquid–solid area and liquid–vapor area normalized by the projected area[37, 38] (see reference[39] for the terminology used here). Assuming minimal liquid penetration into the SHS and that the SHS grains are smooth and devoid of reentrant geometries[34, 40], the predicted apparent (macroscopic) contact angle for $\phi_{LS} = 0.1$ and $\phi_{LV} = 0.9$ yields $\theta_{A,p} = 158°$, which is in reasonable agreement with the experimental observation of $\theta_A \approx 160°$. This means that when a water droplet is placed on an SHS mulch, it is practically hovering on air (because only 10% of its area touches the solid). This also underlies the effortless de-pinning of water from SHS (or ultralow contact angle hysteresis of water).

Next, we explain how SHS governs the fate of water loss via evaporation from the topsoil. First, we emphasize that all the field-scale experiments reported in this study utilized subsurface irrigation. When water comes in contact with common hydrophilic soil particles, the mechanical equilibrium between the interfacial tensions creates a concave meniscus, which drives it in all directions, including upwards, due to capillarity (Figs. 7A-B). Considering that the particle size of soil ranges from 1 to 1000 μm (Fig. 2B), pore sizes between the grains could yield an average radius of curvature of the water meniscus of $r_c \approx 5$ μm. For an actual contact angle of $\theta_o \approx 39°$ for silica, the major component of our soil (please refer to Supplementary Information, Section III, Fig. S15), the magnitude of the Laplace or the curvature pressure can be determined by the formula: $P_L = \gamma_{LV} \times C_m$, where $\gamma$ is the surface tension of water, $C_m =$ $(1/R_1 + 1/R_2)$ is the mean curvature of the liquid–air interface, and $R_1$ and $R_2$ are the mutually orthogonal radii of the interfacial curvature. Assuming spherical symmetry of the air–water interface, $R_1 = R_2 = r_c \approx 5\ \mu m$, which yields $P_L \approx 29\ kPa$. This pressure drives the water radially from the subsurface dripping point in all directions, including upwards to the topsoil–air interface, where it evaporates (Figs. 7A and 1A).

Conversely, when soil moisture rises upward and touches the SHS mulch, the curvature of the air–water interface at the interface of the SHS grains becomes convex (Fig. 7C–D). Therefore, the same Laplace pressure that drives the capillary rise of water prevents its imbibition into the SHS layer; thus, keeping it dry (Fig. 7D). In fact, the magnitude of this preventive (or negative) Laplace pressure can be calculated using the formula $P_L = \gamma_{LV} \times C_m$. We presume that the average value of the $C_m$ of the air–water interface in contact with SHS grains of sizes ranging from 100–700 μm is $\approx 1/60\ \mu m^{-1}$, which yields $P_L = 1.2\ kPa$, a result in reasonable agreement with that of breakthrough pressure experiments. This analysis explains how an SHS layer facilitates a dry porous barrier for water vapor (and other gases).

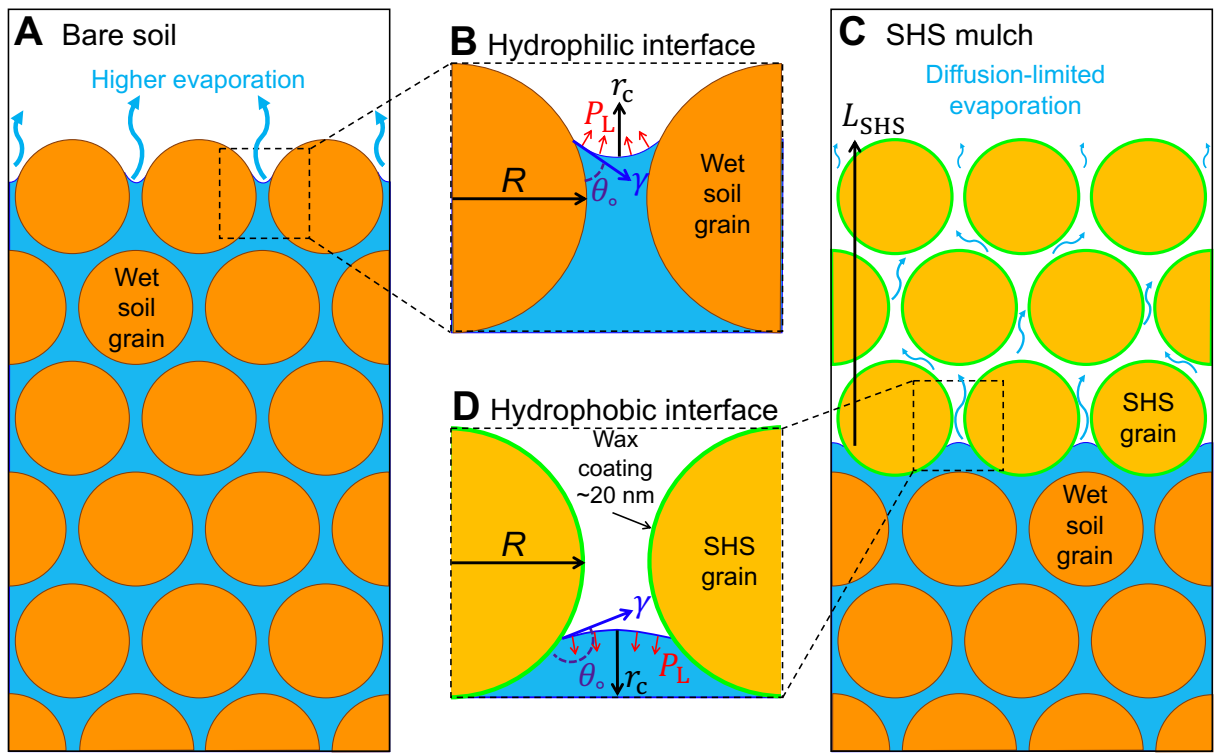

**Fig. 7 Mechanism for limiting water evaporation from the soil.** (**A**) Water from subsurface irrigation is spontaneously imbibed by the soil media due to (**B**) positive Laplace pressure, $P_L$

(red arrows). This results in the capillary rise of water, leading to evaporation loss. (**C**) Water from subsurface irrigation is imbibed by the soil, but this imbibition is arrested at the soil–SHS interface due to (**D**) negative Laplace pressure. Subsequently, SHS acts as a barrier limiting dry diffusion, significantly reducing water loss from the soil.

While liquid water is unable to spontaneously imbibe into SHS, water vapor can diffuse through the SHS mulch layer. In fact, the evaporation flux of water from the topsoil and across the SHS mulch can be estimated using the Fick's diffusion model $J = -D \times \Delta\varphi/L_{\text{SHS}}$, where $\Delta\varphi$ is the gradient in the water vapor concentration across the SHS layer, $L_{\text{SHS}}$, and $D = \frac{\mathfrak{D}_w \times \varepsilon \times \delta}{\tau}$ is the effective diffusivity that depends on the diffusivity of water vapor in air, $\mathfrak{D}_w$, porosity, $\varepsilon$, constrictivity, $\delta$, and tortuosity, $\tau$, of the granular mulch[41]. As evident from this equation, the thicker the SHS mulch, the lower the evaporation flux, as observed by the $\sim\frac{1}{2}\times$ scaling of the evaporation fluxes with the doubling of the SHS mulch thickness (Fig. 3G). SHS also decreases $\Delta\varphi$ by physically covering the topsoil, which insulates the soil-moisture underneath from sunlight and wind.

During the course of the trials, the field station experienced dust storms and daily wind speeds of ~0–15 ms$^{-1}$ without noticeable loss of SHS or its water-conserving properties[42]. This was due to the relatively larger size and mass of the SHS relative to soil particles (Figs. 2B) as well as due to the interparticle friction between SHS grains (Fig. S2F). During the later stages of the crop cycle (March–April), soil temperatures in our fields were more than 70°C. However, this did not compromise the wax coating (bulk melting point ~60°C–65°C); the adhesive force at the sand–wax interface exceeded the weight of the wax, preempting its dripping[43].

**Conclusion**

Taken together, the results presented in this study underscore the potential of our bio-inspired approach for boosting food production in arid regions. Among its ingredients, paraffin wax is the most inexpensive hydrophobic coating available at an industrial scale, in contrast with costly organic/perfluorinated silanes[14, 44]. Furthermore, sand can be acquired directly from sandy soils, which are common in arid regions, such as the Middle East. For example, the soil in our field station comprised ~80% sand (Figs. 2B, S15, and SI Section III). Presently, we are developing a solvent-free manufacturing protocol that directly exploits sandy soils and paraffin wax to produce SHS. The resulting SHS also exhibits excellent water repellency (Fig. S3) and its field trials are underway.

To summarize, this translational research demonstrates how sands or sandy soils – abundant resources in arid regions – can be processed into SHS via nanoscale wax coating to produce more food with limited water resources. The simplicity of this manufacturing process is underscored by the fact that we manufactured over 10,000 kg of SHS in our laboratory and that we could manually apply it to conduct these field trials (Movie S2). In addition to annual crops, we believe that SHS can be applied to perennial orchards, vineyards, or green zones in cities, such as public parks and green corridors. Arid lands with limited freshwater sources, including parts of South America, Africa, the Middle East, the United States, Australia, China, and India, present vast underused regions with untapped potential to expand their agricultural operations by leveraging this sustainable technology.

**Methods**

**Process of manufacturing SHS**

1. Starting materials:
    a. Common sand (characteristic size in the range: 100–700 μm; Fig. 2B).
    b. Blocks of paraffin wax (molecular weight, 487 Da, melting point 60–65°C), grated to shavings of <1-mm size.
2. These wax shavings were dissolved in approximately 10 L of hexane.
3. Then, 50 kg of common sand was added to a 60-L evaporator (Fig. S1).
4. The wax/hexane mixture was added to the evaporator flask to achieve a wax-to-sand ratio of 1:600. Note that concentrations as low as 1:2000 (wax/sand) achieved superhydrophobicity.
5. Next, the temperature of the evaporator was gradually increased from 22°C to 55°C and the pressure was reduced from 1 atm to 100 mbar to evaporate the hexane, which was condensed and collected for reuse (recovery rate of ~99%).
6. The pressure was normalized and SHS was collected and stored for use.

Note: Due to the flammability of the organic solvents used in this manufacturing protocol, potential sources of electrical sparks, such as cell phones, match sticks, and static charge, must be avoided and the setup must be installed in a well-ventilated area. All components of the setup were electrically grounded to avoid any electric spark.

More recent developments have shown that the solvent can be completely eliminated and replaced by vigorous mixing at 70°C–80°C for approximately 30 min without noticeably compromising the coating.

**SHS and wax characterization**

The wax used was characterized via gas chromatography–mass spectroscopy (Agilent 7890A–5975C) (~$10^{-1}$ M in cyclohexane). The wetting of individual particles was observed under an environmental scanning electron micrograph (FEI Quanta 600). Contact angles on a 10-mm bed of SHS were measured by placing a 10-μL water droplet with advancing and receding rates of 0.2 μL/s (Kruss Drop Shape Analyzer DSA100, Advance software). The thickness of the wax layer on the sand granules was estimated from the surface area of sand grains and the volume of wax used. The surface area of the sand was estimated using the Brunauer–Emmet–Teller (BET) method using krypton gas[45]. We found it to be 0.11 m$^2$/g. For 1 g of wax of mass density 800 kg/m³, the volume was 1.25 ×10⁻⁶ m³, whereas for 600 g of sand, the surface area was 600 g × 0.11 m²/g = 66 m². Assuming uniform coating, the average thickness of the wax layer was 1.25 × 10⁻⁶ m³/66 m² = 1.8×10⁻⁸ m ≈ 20 nm). More details are presented in Section I of the Supplementary Information.

**Field trials with tomato, barley, and wheat**

Field trials were conducted at the agricultural research station of King Abdulaziz University, Hada Al Sham, Saudi Arabia (21.7963° N, 39.7265° E). Tomato (*S. lycopersicum*) varieties A (Bushra) and B (Nunhem's Tristar F1), wheat (*T. aestivum*) variety Balady, and barley (*Hordeum vulgare*) variety Morex. The field was divided into plots of 2.5 × 2.5 m for each treatment, with plants uniformly spaced within each plot. The plots were evenly distributed over the field (Fig. 5C). The total number of tomato plants per plot was 12 and 15 in the years 2018 and 2019, respectively. For barley and wheat, the seeds were spread in a line 2.5 m long and 5 cm deep, with 15-cm lines in between. The distance between the plants was adjusted at approximately 2 cm by thinning them after full germination. To control the thickness of the SHS mulch, we used a cardboard template of known area and homogeneously applied a specific mass of SHS (Movie S2). Tomato fields received SHS mulch as a 40-cm strip around the plants in a line for the 2018 season and as 40-cm diameter circles around each plant for the 2019 season to save on material. On the other hand, barley and wheat fields were completely covered with the SHS mulch due to the close proximity between plants. Plastic mulches (120-μm thick polyethylene sheets) were also applied over the whole plot. For the subsurface drip irrigation system, we used Rain Bird LD-06-12-1000 Landscape drip 0.9 Gallon/h (3.4 L/h per dripping point). The plants were fertilized through the irrigation system using common compound fertilizers on a weekly basis (N:P:K 20:20:20 during the vegetative stage and N:P:K 10:10:40 at the flowering and fruiting stages). For the 2019 season, there were two plots with 5-mm thick SHS mulch, clear plastic (transparent 120-μm thick polyethylene sheet), and black plastic (120-μm thick polyethylene sheet) and four plots with 10-mm thick SHS and unmulched controls. The number of plots and replicates (*n*) varied according to the treatment, as evidenced by the number of dots in Fig. 5A–B. More details are presented in Section II of the Supplementary Information.

Tomato fruits were harvested weekly and collected in plastic bags labeled with plant identity number, followed by counting and weighing. Weekly harvests were conducted for 6 weeks, and the total number and mass of fruits per plant was considered for data analysis (Fig. 5A). Barley and wheat plants were collected at the end of the cycle and counted over a homogeneous area of 1 m² inside each plot. The plants were manually separated and counted, followed by extrapolating the yield per plant to yield per hectare by multiplying the total number of plants in the measured area by the mass of grains per plant (Fig. 5B). Extrapolation to hectare was preferred to avoid the inherent difficulty in separating one plant from another due to their close proximity and tangled roots. Statistical analysis of the treatments was performed in Matlab R2019b using the Kruskal–Wallis H test, where (S) in Fig. 5 represents statistical significance ($p < 0.05$). In addition, we performed soil analysis for all plots to guarantee homogeneity within the field; the details are presented in Section III of the Supplementary Information.

**Quantification of evaporation flux**

Pots containing 75 g of Metro Mix 360™ soil and 132 g of water were covered with either SHS mulch (0-, 5-, 10-, 15-, and 20-mm thick) or common (uncoated) sand (0-, 5-, 10-, 15-, and 20-mm thick) or were not subjected to any mulching (bare) and placed outside the lab under environmental conditions (Fig. 3). All pots were in duplicates. The masses of individual pots were tracked over time with a mass balance. The experiment was started on October 10, 2016 at 9:45 a.m. at Thuwal, Saudi Arabia (22.3084° N, 39.1030° E). The environmental data were obtained from KAUST weather station.

**Quantification of soil moisture**

We buried 30 L plastic buckets with a diameter of 35 cm at the ground level and filled them up with local sandy soil (Fig. 4A-C), which was sieved to remove gravel of size >1 cm. We compared the 5-mm thick SHS mulch with the unmulched control. Bottom-perforated buckets were used to circumvent the issue of lateral percolation as well as to measure vertical flows, i.e., evaporation and percolation. To quantify soil moisture and temperature, we used Hydraprobe™ sensors (Stevens Water, LLC), which exploit the principle of coaxial impedance dielectric reflectometry[46, 47]. Briefly, we compared mulched and unmulched soil under two irrigation regimes: (i) single irrigation starting with supersaturated soil with no further irrigation and (ii) daily subsurface drip irrigation (1.1 L/day). The data for case (i) were determined as the average measurement of two sensors buried at a depth of 5 cm (Fig. 4D). The data for case (ii) were estimated as the average measurement of three sensors at depths of 15, 20, and 25 cm (Fig. 4E). The experiments were performed from January to April (2018) at the same site as our field trials (21.79° N, 39.72° E). The weather conditions of the site are presented in Fig. S4.

**Analysis of soil–root microbiome**

Samples were collected during the 2018 crop season on February 20 at the same site as our field trials (21.79° N, 39.72° E). For each field treatment (irrigation: fresh/brackish water and overlay: SHS mulches presence/absence), the root systems of barley and tomato plants were randomly selected from each experimental plot (barley, $n = 18$; tomato, $n = 18$; Table S5). After gently removing the plants from the soil, the root system was sampled using a pair of sterile scissors and tweezers. The rhizosphere was separated from the root tissues, as described previously[48]. The obtained root tissues were surface-sterilized according to a previously reported methodology[49]. All samples were stored at −20°C for molecular analysis. Bulk soil samples were also collected from the unvegetated portions of each experimental plot (0.5–10-cm depth; $n = 18$ for each of the two crops).

DNA was extracted from $0.5 \pm 0.05$ g of bulk and rhizosphere soils using the PowerSoil® DNA Isolation kit (MoBio Inc., USA). Surface-sterilized root tissues were ground in a mortar and pestle using liquid nitrogen, followed by DNA extraction from 1 gram of the ground root tissues using the DNeasy Plant Maxi kit (Qiagen, Germany). The V3–V4 hypervariable regions of the 16S rRNA gene were PCR-amplified using the universal primers 341F and 785R[50]. Libraries were constructed using the 96 Nextera XT Index Kit (Illumina) following the manufacturer's instructions and sequenced using the Illumina MiSeq platform at the Bioscience Core Lab at KAUST. Raw sequences were analyzed using the DADA2 pipeline, including quality filtering, trimming, dereplication, and paired-end merging of the sequences[51]. Chloroplast- and mitochondria-classified SVs were discarded and non-prevalent SVs (defined as SV present in <1% of our samples) were removed from the SV table. A total of 13,568,983 sequences (barley: 8,611,818 and tomato: 4,194,505) divided into 6,912 SVs were obtained. Only samples with suitable sequencing depth and diversity (Good's coverage value > 99%) were used for the further analysis. The raw reads have been deposited in the Short Reads Archive of NCBI under the accession numbers SUB7136748, SUB7211346, and SUB7137862.

The beta-diversity of the bacterial communities in barley and tomato was analyzed using the compositional similarity matrices (Bray–Curtis) of the relative log-transformed SV tables using a previously reported methodology[52]. The same matrices were also used to perform unconstrained analysis of principal coordinates and to assess the compositional variation explained by each experimental factor. Permutational multivariate analyses of variance (PERMANOVA) were performed using the *adonis* function from the "vegan" package[53] to statistically test the impact of each experimental factor: "irrigation" (two levels: fresh and brackish water), "overlay" (two levels: SHS mulch and bare soil), and "compartment" (three levels: root tissues, rhizosphere, and bulk soil). Further, PERMANOVA pair-wise tests were conducted to evaluate the effect of "overlay" on each interaction category "Compartment" × "Irrigation" using PRIMER[52].

**Data availability:** The authors declare that all the data supporting the findings of this study are available within the paper and in its Supplementary Information.

**Acknowledgments:** We thank the following colleagues from KAUST: Ms. Jamilya Nauruzbayeva for her assistance in applying SHS in the fields; Mr. Vinicius Luis dos Santos for his assistance in conducting soil analysis; Dr. Mahmoud Ibrahim and Dr. Andreia Farinha for troubleshooting during soil analysis; Prof. Simon Krattinger for data analysis; Mr. Sankara Arunachalam and Dr. Eddy Domingues for environmental scanning electron microscopy and energy-dispersive spectroscopy; Dr. Erqiang Li and Professor Sigurdur Thoroddsen for high-speed imaging; Dr. Yoann Malbeteau and Mr. Bruno Aragon for their assistance in soil moisture data analysis; Dr. Nishan Abdul Jaleel, Dr. Gabriele Fiene, and Dr. Muppala Reddy for their assistance in greenhouse experiments. We would also like to thank Dr. Adel D. Al-Qurashi, Prof. Dr. Abdullah S. Al-Wagdany, and Dr. Khalid Asseri from KAU, Jeddah. The co-authors thank Mr. Xavier Pita, Scientific Illustrator at KAUST, for creating Fig. 1A-B and Dr. Michael Cusack (KAUST) and Prof. Kevin Plaxco (University of California, Santa Barbara) for their assistance in scientific editing.

**Funding:** We acknowledge research support from King Abdullah University of Science and Technology (KAUST) through award # BAS/1/1070-0101 to HM.

**Author contributions:** HM conceived the idea of superhydrophobic sand, supervised this research, and coordinated the collaboration; AGJ scaled-up the SHS manufacturing protocol and led field-scale experiments with the assistance of MAAM, KO, and EM; JR performed initial feasibility studies with SHS, assisted by MJLM, MT, and MAAM; AGJ and SM performed the soil-moisture measurements and MFM advised them on sensor calibration and data analysis; MAAM and MT provided expert guidance on field-scale experiments; DD, RM, and KO performed microbial analysis and analyzed the data; AGJ and HM analyzed the data from the field-scale experiments; HM and AGJ wrote the manuscript and the co-authors edited it.

**Competing interests:** HM, AGJ, and JR have filed a patent application WO2018091986A1.

**Data and materials:** Raw data have been deposited in the Short Reads Archive of NCBI (Accession numbers: rhizosphere, SUB7136748; root tissue, SUB7211346; and bulk soil, SUB7137862); remaining data are available in the main text and supporting information. Code is available upon request.

# Supplementary Information

## Superhydrophobic Sand Mulches Increase Agricultural Productivity in Arid Regions

Adair Gallo Jr.[1], Kennedy Odokonyero[1], Magdi A. A. Mousa[2,3], Joel Reihmer[1], Samir Al-Mashharawi[1], Ramona Marasco[4], Edelberto Manalastas[1], Mitchell J. L. Morton[4], Daniele Daffonchio[4], Matthew F. McCabe[1], Mark Tester[4], Himanshu Mishra[1*]

[1]King Abdullah University of Science and Technology, Water Desalination and Reuse Center, Division of Biological and Environmental Sciences and Engineering, Thuwal 23955 - 6900 Saudi Arabia

[2]Department of Arid Land Agriculture, Faculty of Meteorology, Environment and Arid Land Agriculture, King Abdulaziz University, 80208 Jeddah, Saudi Arabia

[3]Department of Vegetables, Faculty of Agriculture, Assiut University, 71526 Assiut, Egypt

[4]King Abdullah University of Science and Technology, Division of Biological and Environmental Sciences and Engineering, Thuwal 23955 - 6900 Saudi Arabia

[*]Corresponding Author: himanshu.mishra@kaust.edu.sa

**Content:**
Supplementary Text
Figs. S1 to S16
Tables S1 to S6
Captions for Movie S1 and S2

    SECTION I – MATERIALS CHARACTERIZATION
    SECTION II – FIELD TRIALS
    SECTION III – FIELD SOIL CHARACTERIZATION

**Additional files:**
Movie S1
Movie S2

# SECTION I – MATERIALS CHARACTERIZATION

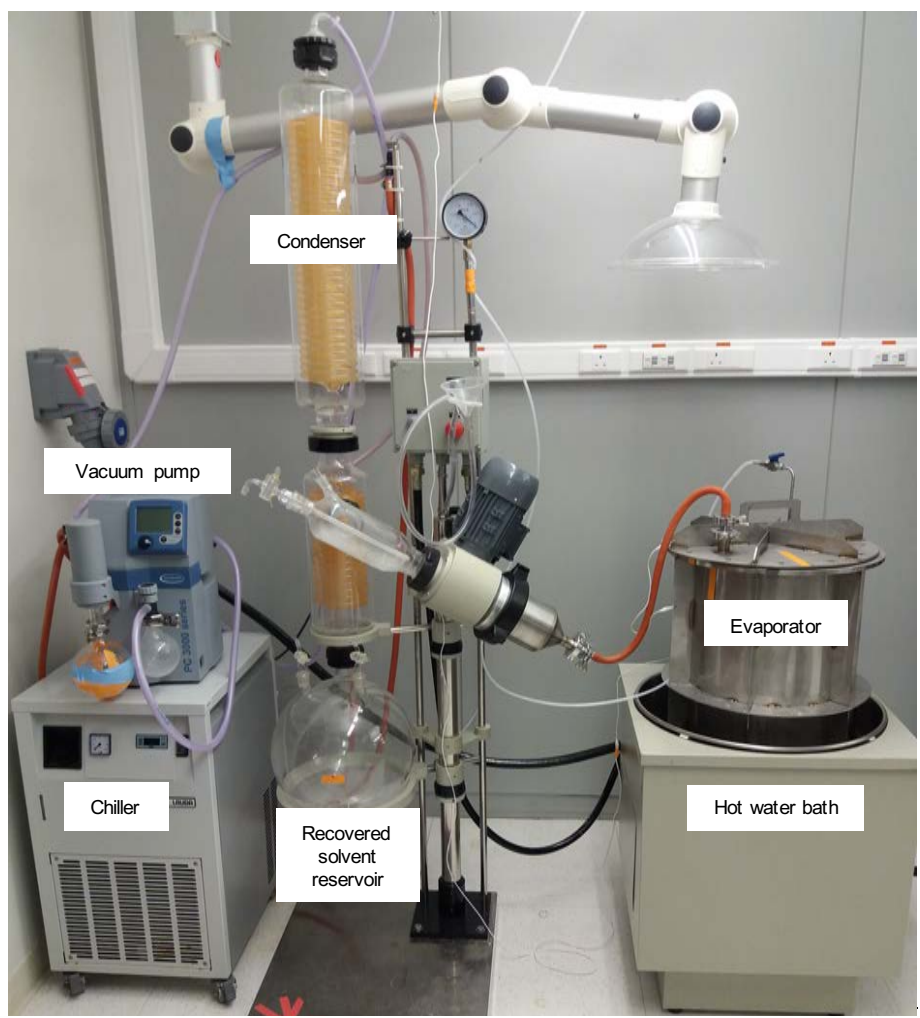

**Fig. S1 – Photograph of the superhydrophobic sand (SHS) manufacturing setup.** Sand was loaded into the evaporator flask (~50 kg) along with a solution of paraffin wax (1:600 wax to sand ratio) in hexane (1 L:5 kg, hexane to sand). The bath temperature was set at 55 °C and the pressure was reduced to 100 mbar to evaporate the hexane, which was condensed and collected for reuse.

We measured the carbon and proton nuclear magnetic resonance ($^{13}$C-NMR, $^{1}$H-NMR) spectra of the paraffin wax in 500 MHz Bruker Automated sampler NMR. For $^{13}$C-NMR and for $^{1}$H-NMR we used 256 scans and 512 scans, respectively. We dissolved a sample of paraffin wax in deuterated chloroform (~$10^{-1}$ M) and analyzed it in 5 mm NMR tubes. In Fig. S2A we present the spectra for $^{13}$C-NMR, showing a major peak around 29–30 ppm and peaks at 22.8 and 32.1, which correspond to carbons from $CH_2$[1]. The less intense peak at 14.1 ppm corresponds to carbon from $CH_3$ [1]. The small signal at 34 ppm could be an indication of CH in some ramifications. The lack of peaks above 40 ppm indicates the absence of double bonds or functional groups, such as acids, esters, or alcohols. In the $^1$H-NMR spectra (Fig. S2B), the strong signal around 1.3 ppm is indicative of H from $CH_2$ and the peak at around 0.9 ppm indicates H from $CH_3$ [2]. The deuterated chloroform peak appears at 7.25 ppm. The NMR data thus confirms the wax as mostly a mixture of linear-chain alkanes without functional groups.

Using energy dispersive spectroscopy (EDS), coupled to the SEM, we obtained a semi-quantitative analysis of the chemical composition of the surface of our sand granules. The results demonstrated that the SHS surface was, as expected, richer in carbon content in comparison to that of ordinary sand. Specifically, the atomic percentage (At.%) of carbon in normal sand was 3.3% (organic contaminants)[3] while in SHS it was 19.5% (Fig. S2C–D). The increase in the characteristic X-ray signal of carbon and corresponding decrease in the silicon signal demonstrate an addition of hydrocarbons to the surface of the sand granules. A small peak of gold is observed because we coated a 4 nm thick gold layer to prevent electrical charging of our (otherwise insulating) samples under electron radiation. As expected, the major peaks are for $SiO_2$ because the penetration depth of EDS is greater than the thickness of the wax coating.

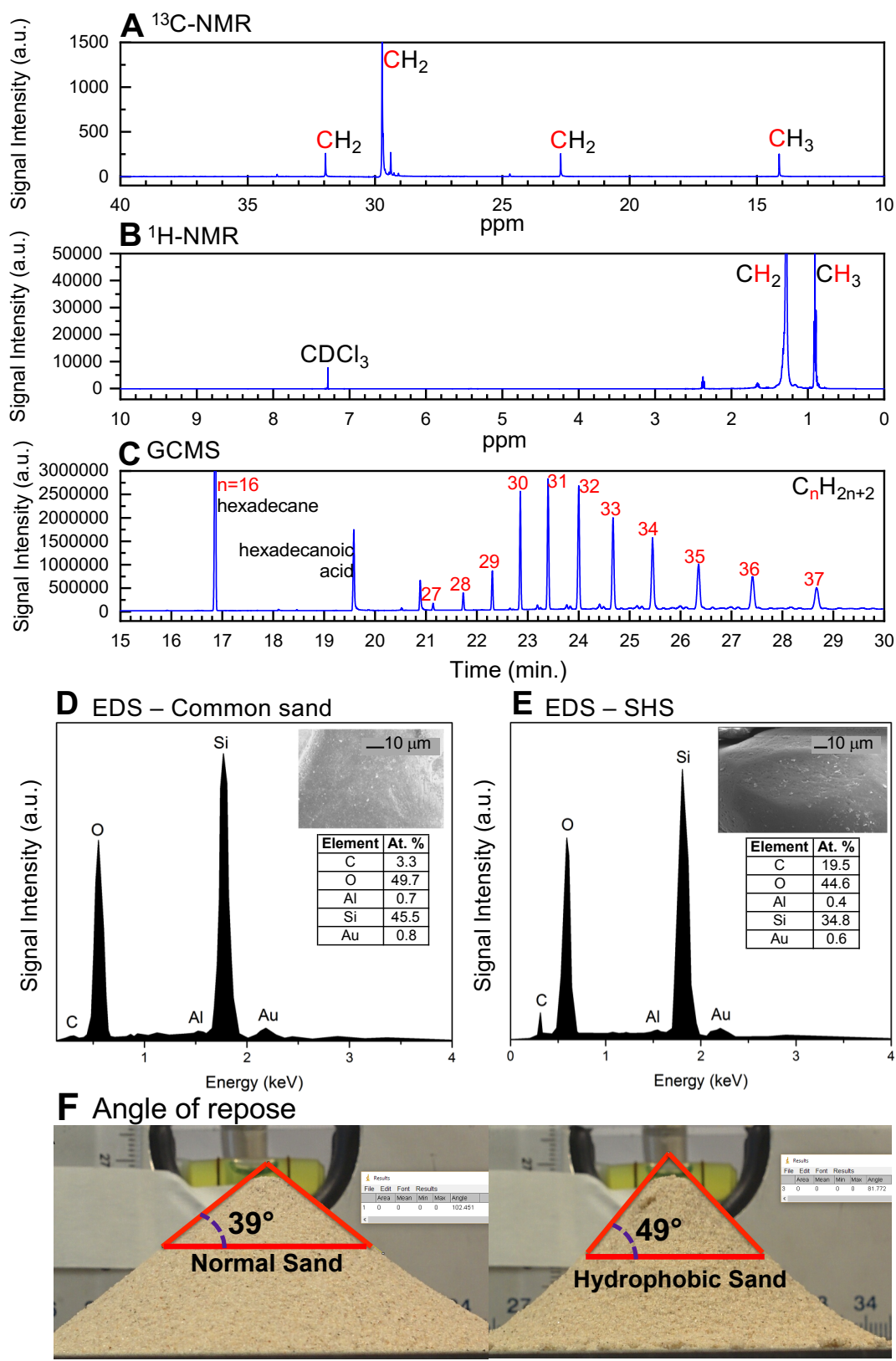

**Fig. S.** ... **SHS.** (**A**) Carbon and (**B**) proton nuclea... a of the paraffin wax (~$10^{-1}$ M in deuter... matogram (GC) of the paraffin wax (~$10^{-1}$ ... (n) of each alkane in red. Energy disper... mmon uncoated sand and (**E**)

superhydrophobic sand (SHS) with SEM images (top right insets). (**F**) Angles of repose of uncoated sand versus SHS.

Additionally, we tested the influence of temperatures above the melting point of wax (60—65 °C). We heated the SHS up to 120 °C for 1 hour and let it cool down back to ambient temperature (23 °C). Next, we conducted typical contact angle measurements and we did not observe any significant effect on the coating or to the superhydrophobic properties of the SHS.

The addition of paraffin wax to the surface of the sand grains slightly increased the friction between then, evidenced by the decrease in the angle of response (cone angle) from ~38.5° to ~49°.

We performed the gas chromatography mass spectrometry (GCMS) in a 7890A Agilent Gas Chromatograph. equiped with DB-5ms GC column (30 m, 0.25 µm). We dissolved the wax sample in cyclohexane (~$10^{-1}$ M) and analyzed it with a temperature ramp of 50 °C to 120 °C, at 15 °C/min, then holding for 5 min., then up to 320 °C, at 15 °C/min, then holding for 7 min. We used hexadecane ($n$ = 16) as a reference point. We fitted a quadratic equation to the number of carbons in the alkane as a function of the retention time and then aligned the peaks with the known peak for hexadecane (Fig. 2B). In Table S1 we show the concentration of the alkanes in the wax, with 16% of $C_{32}H_{66}$, the alkane in greatest concentration in the wax.

**Table. S1 – Chemical composition of the paraffin wax used in our experiments, obtained via GCMS.** The general formula for the linear-chain alkanes is $C_nH_{2n+2}$.

| Retention time (min.) | Number of carbons in the alkane ($n$) | Molar fraction |
|---|---|---|
| 21.2 | 27 | 0.7% |
| 21.8 | 28 | 1.9% |
| 22.3 | 29 | 4.3% |
| 22.9 | 30 | 12.5% |
| 23.4 | 31 | 15.1% |
| 24.0 | 32 | 16.3% |
| 24.7 | 33 | 14.0% |
| 25.5 | 34 | 13.1% |
| 26.4 | 35 | 10.1% |
| 27.4 | 36 | 8.4% |
| 28.7 | 37 | 3.6% |

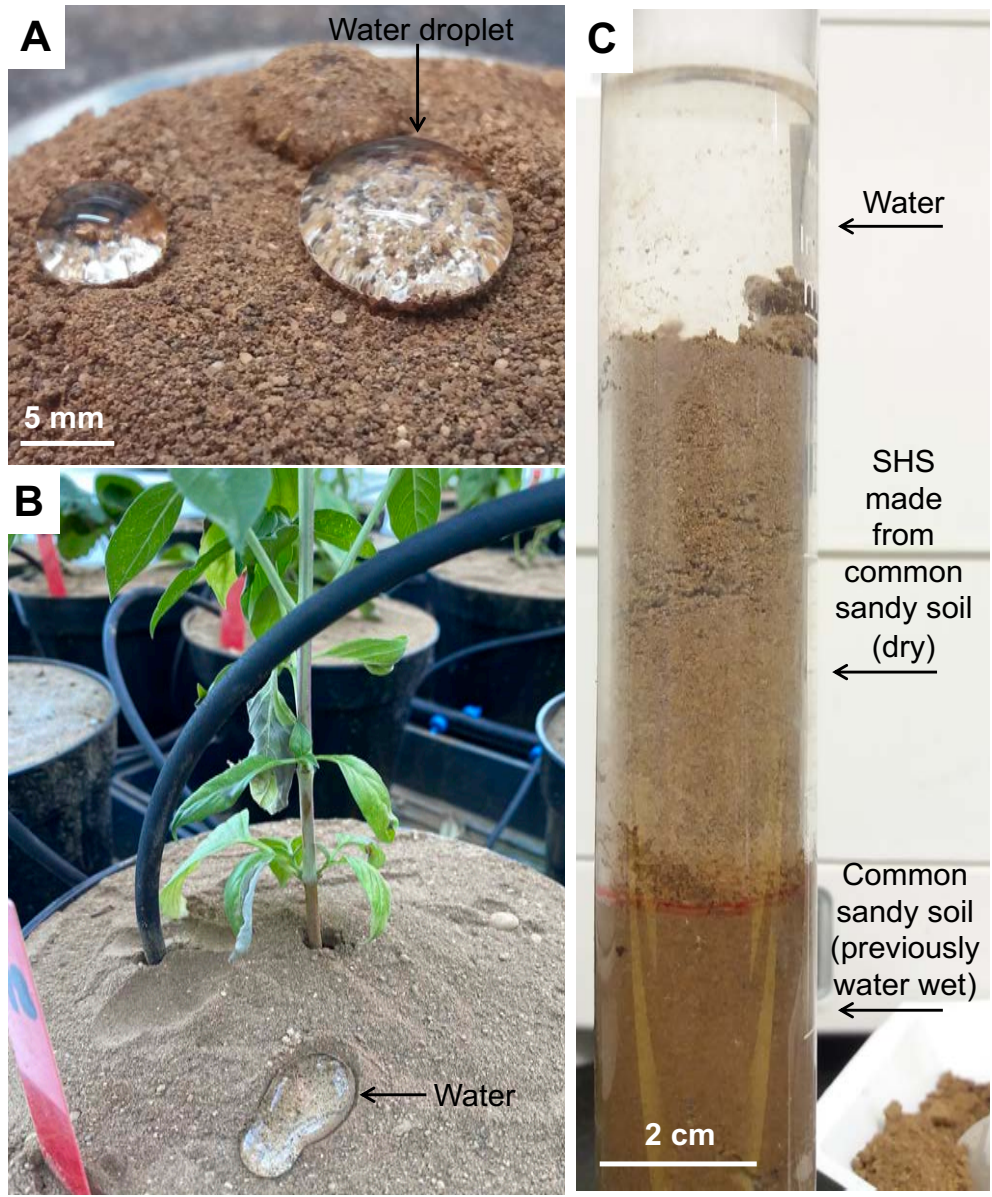

**Fig. S3 – SHS made from local loamy sand soil with the solvent-free method.** (**A**) SHS made from local loamy sand soil previously sieved to partially remove smaller silt and clay fractions. (**B**) SHS mulch made entirely from loamy sand soil. (**C**) SHS made entirely from loamy sand soil separating a wet uncoated soil layer from a top water column. Soil collected from field station in *Hada Al Sham*, Saudi Arabia (21.7963° N, 39.7265° E).

# SECTION II – FIELD TRIALS

We conducted field trials at KAU agriculture research station, *Hada Al Sham*, Saudi Arabia (21.7963° N, 39.7265° E). The region has a loamy sand soil (Fig. 2A) and during the winter it experiences temperatures suitable for crop growth (Figs. S4–S5). The experimental field is situated in a catchment basin which collects rainwater and supplies groundwater for a number of local farms. In Figs. S7–S10 we present photographs of field trials.

For the season of 2018, we divided the field into two sections, one for fresh water irrigation (<900 ppm NaCl) and another for brackish water (~5000 ppm NaCl). For each water treatment, we split the tomato fields into alternating plots of SHS and bare soil (control). Similarly, for barley, for each of the two water treatments we split the barley field into alternating plots of SHS and bare soil (control). We planted tomato seeds in trays and then we transplanted the seedlings to the field after 30 days, while we planted barley directly in the field. We set the irrigation time to 2 periods of 10–20 mins (depending on the plant growth stage and weather)—first at 7 a.m. and then at 5 p.m. In Fig. S4 we present the experimental timeline and weather conditions. Initially we irrigated both fields with fresh water; then we introduced brackish water in the salt treatments simultaneously with the application of SHS mulch (January 18, 2018).

For the season of 2019, we divided the field into two sections: (i) normal fresh water irrigation and (ii) reduced (50%) fresh water irrigation; we irrigated with (i) for 2 periods of 10–20 min at 7 a.m. and at 5 p.m., and with (ii) only in the morning at 7 a.m. for 10–20 min. In Fig. S5 we present the experimental timeline and weather conditions . Initially we irrigated both fields with fresh water twice per day, and then initiated the reduced irrigation treatment simultaneously with the application of SHS and plastic mulches (January 13, 2019).

For all crops we calculated the amount of fertilizer based on the water volume in the tank and the usage rate. We dissolved the fertilizer externally and added it to the tanks filled with about 6000 L of water. We added the compound fertilizers (N:P:K 20:20:20 during vegetative stage and N:P:K 10:10:40 at flowering and fruiting stages) from a local market (2 g/L) on a weekly basis. All treatments and their respective controls were supplied by the same water tanks. We thus ensured that they received the same amount of irrigation and fertilizer.

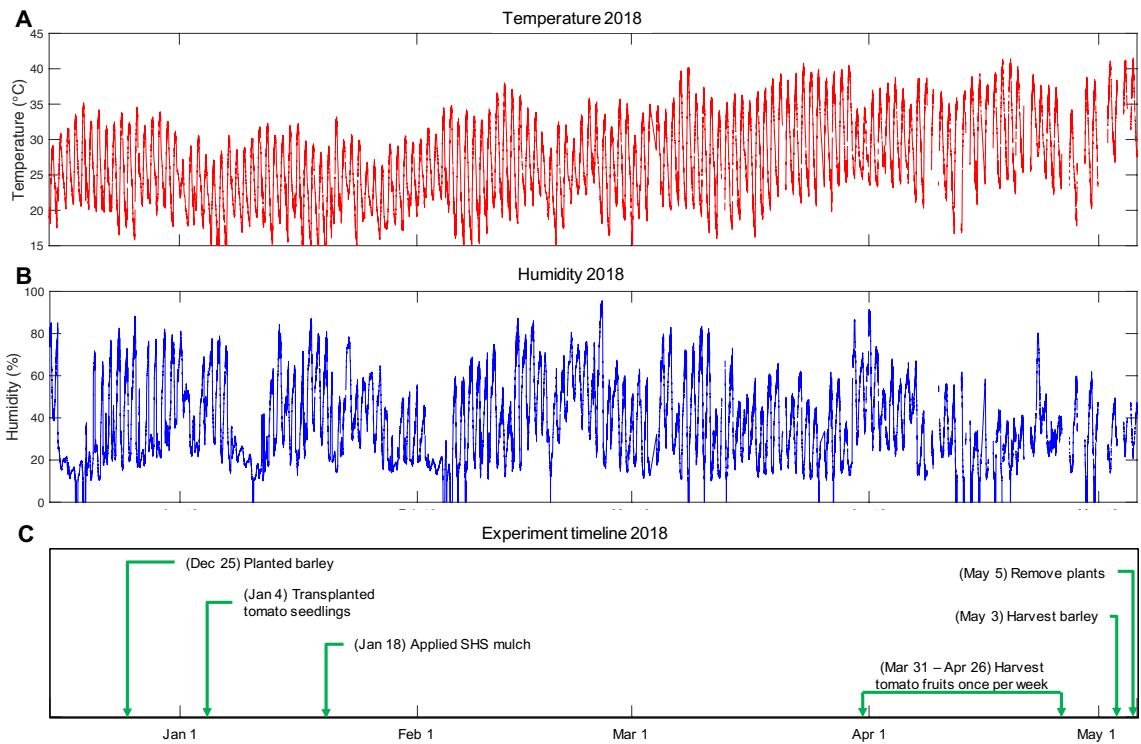

**Fig. S4 – Weather details for tomato and barley field trials (season 2018).** (**A**) Air temperature and (**B**) humidity, and (**C**) experiment timeline. *Hada Al Sham*, Saudi Arabia (21.7963° N, 39.7265° E).

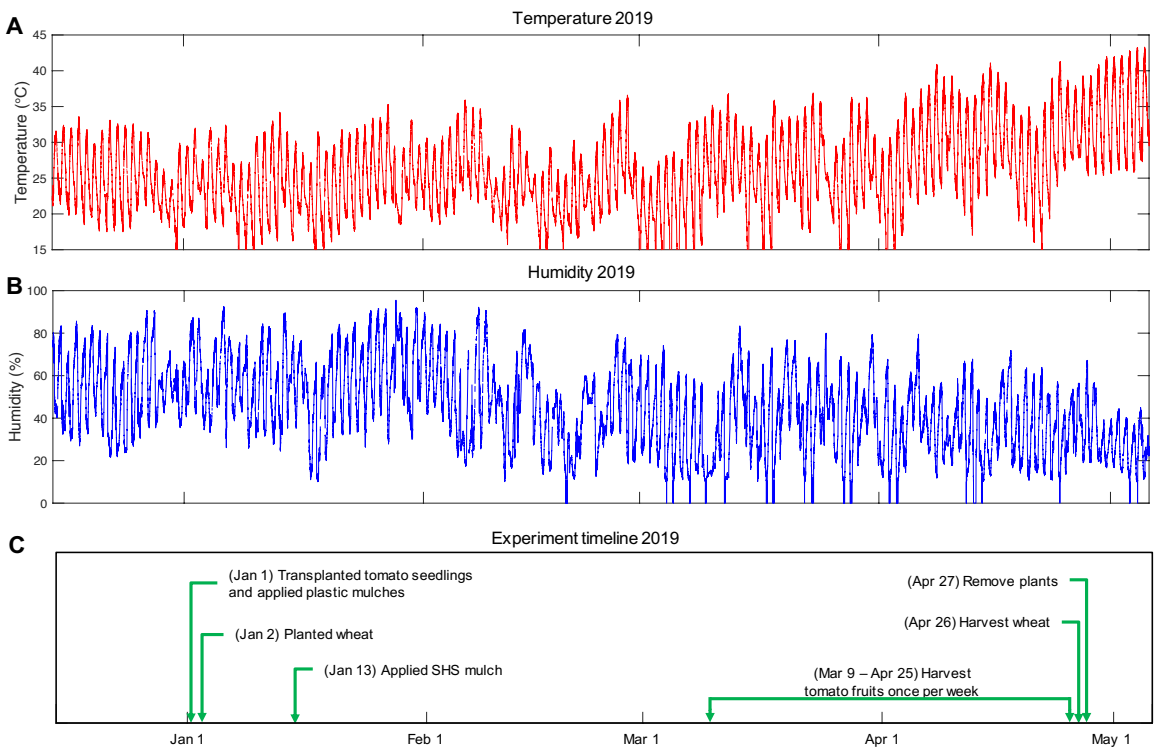

**Fig. S5. Weather details for tomato and wheat field trials (season 2019).** (**A**) Air temperature, (**B**) humidity, and (**C**) experiment timeline. *Hada Al Sham*, Saudi Arabia (21.7963° N, 39.7265° E).

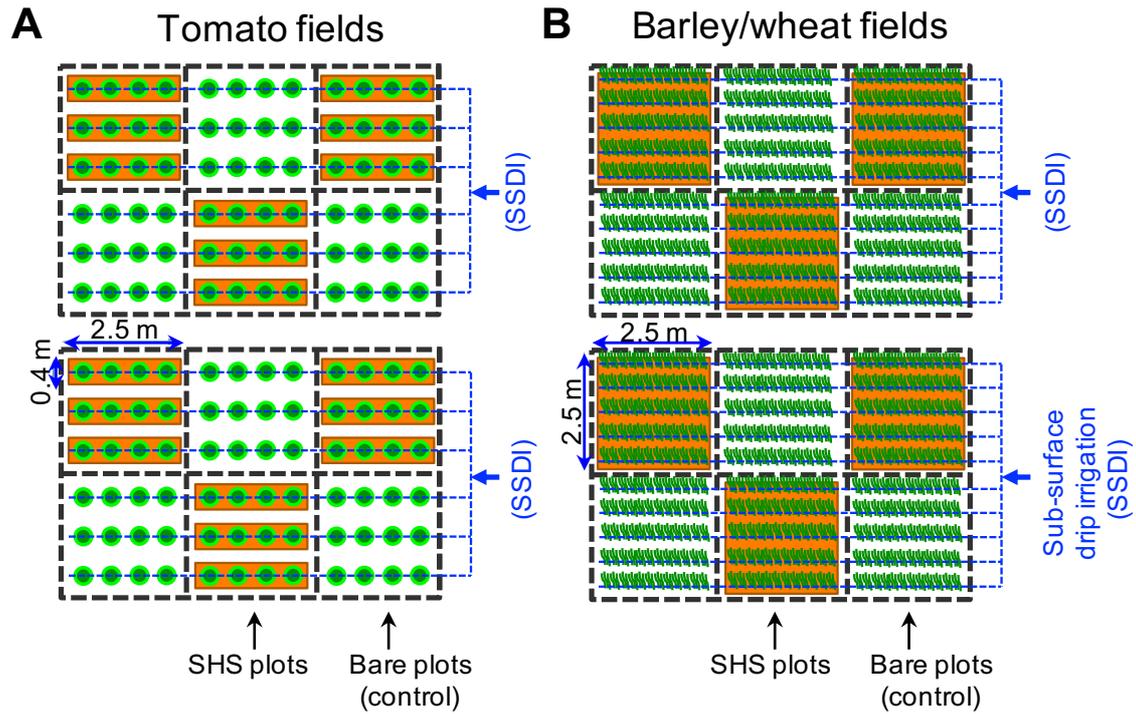

**Fig. S6 – Schematics of field designs.** (**A**) We split tomato and (**B**) barley/wheat fields into 2.5 m × 2.5 m plots with intercalated SHS and bare soil (control). Sub-surface drip irrigation provided equal amounts of water to mulched and unmulched plants. The tomato plots received 3 parallel irrigation lines and the barley and wheat plots received 7 parallel irrigation lines.

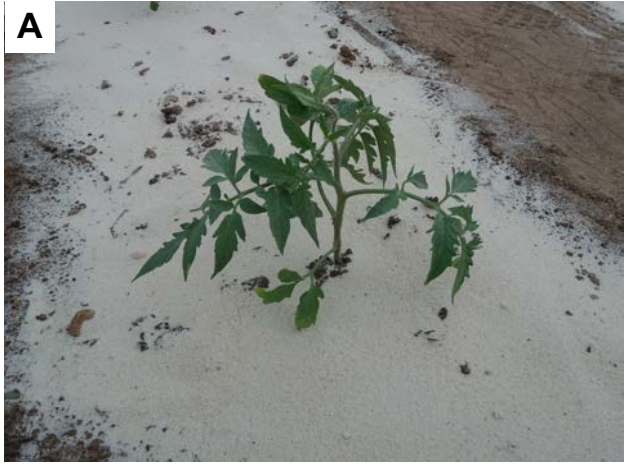 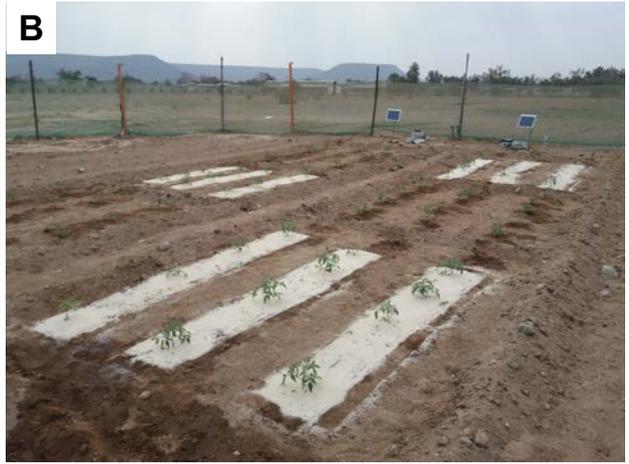
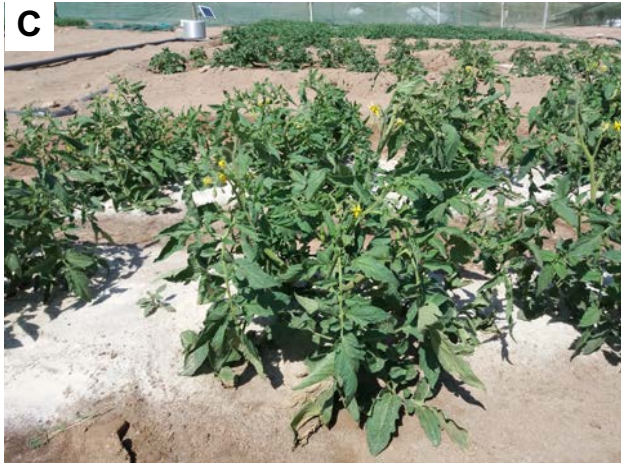 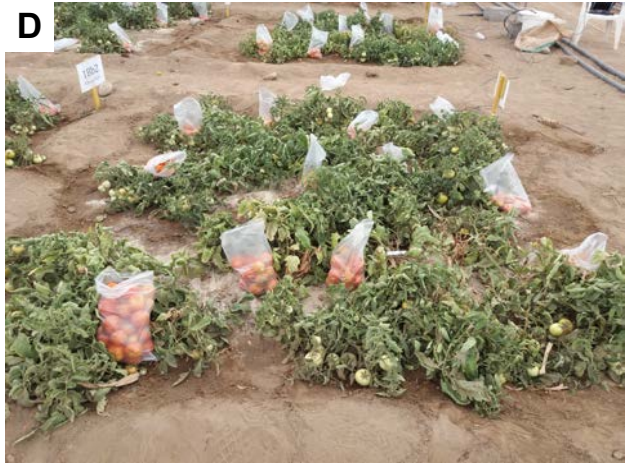

**Fig. S7 – Photographs of the tomato fields (season 2018).** (**A**–**B**) Just after application of the SHS mulch, (**C**) during flowering stage, and (**D**) during harvest.

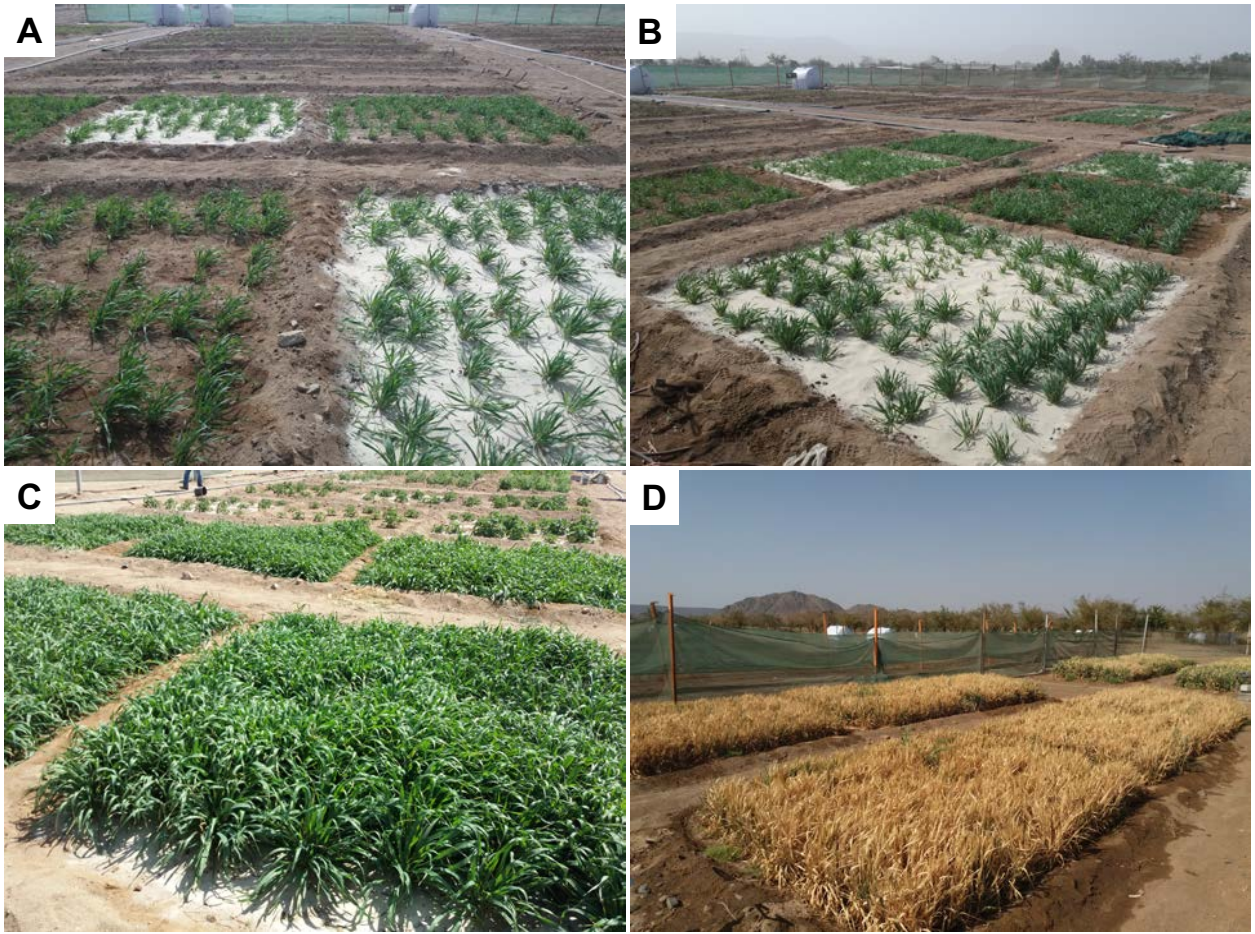

**Fig. S8 – Photographs of the barley fields (season 2018).** (**A–B**) Just after application of the SHS mulch, (**C**) during mid-stage, and (**D**) during harvest.

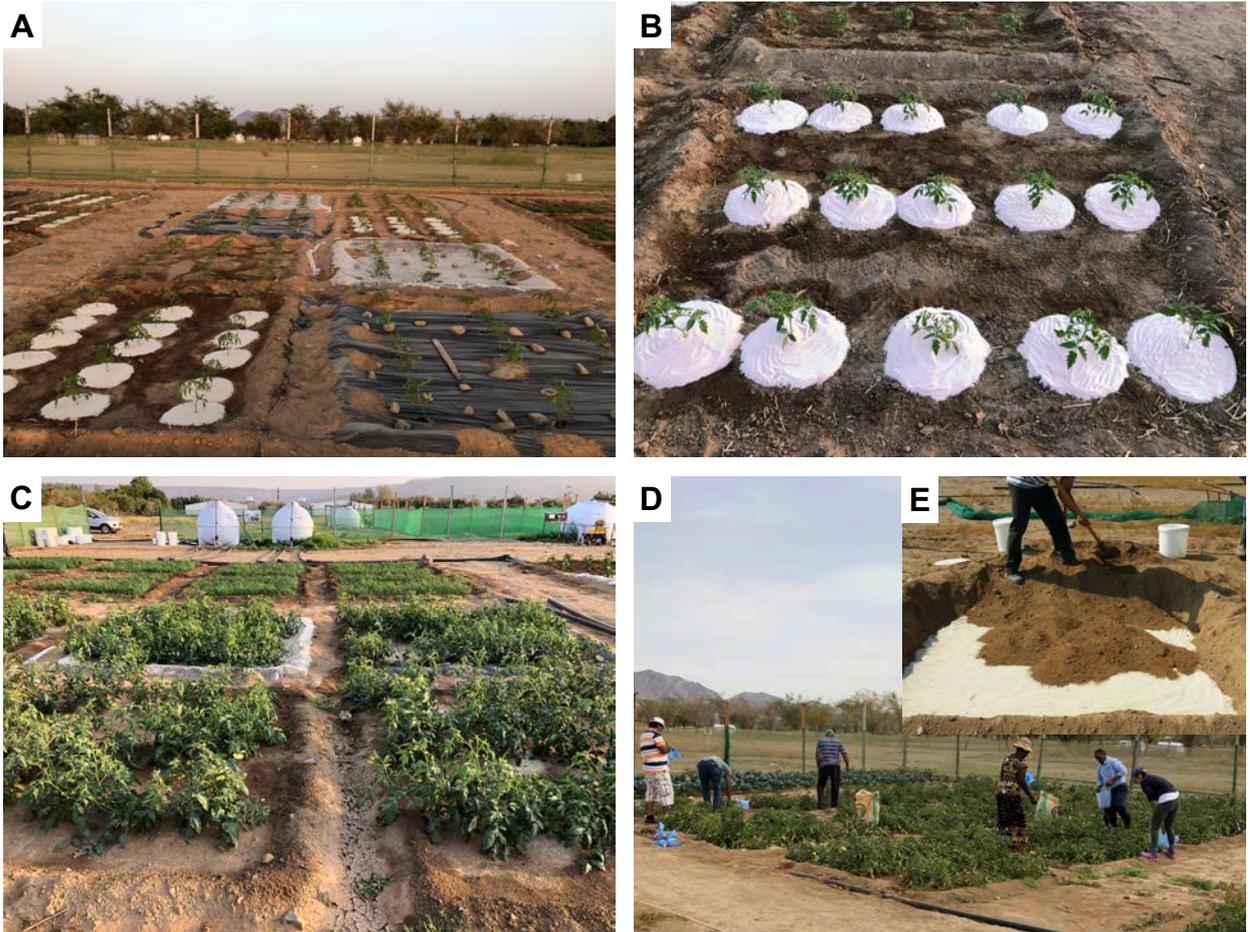

**Fig. S9 – Photographs of the tomato fields (season 2019).** (**A**–**B**) Just after application of the SHS and plastic mulches, (**C**) during flowering stage, and (**D**) during harvest. (**E**) Details of the SHS applied at 40–50 cm depth as a layer to prevent percolation (code: 10/20 SHS).

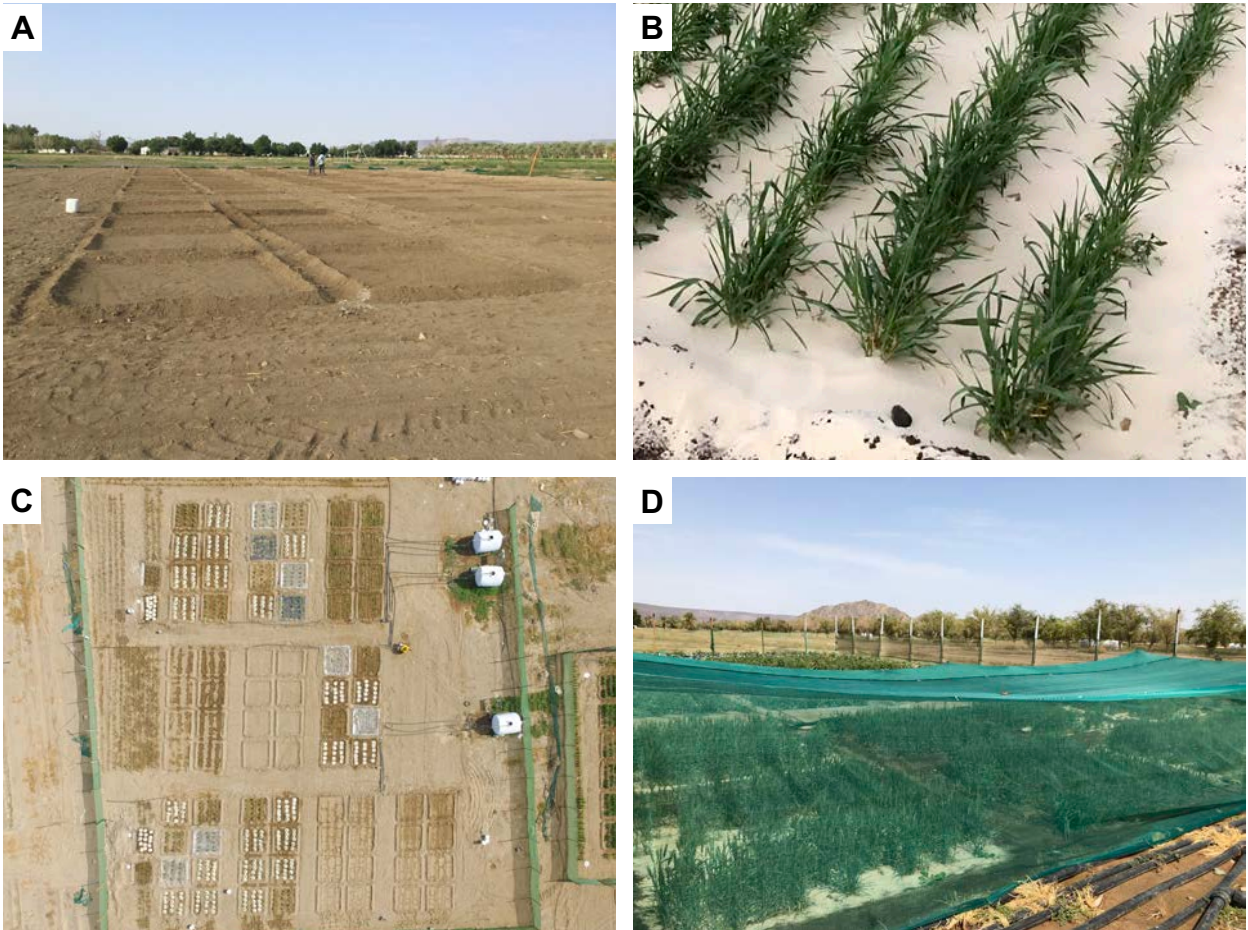

**Fig. S10 – Photographs of the wheat fields (season 2019).** (**A**) Field preparation, (**B**) detail of the application of the SHS mulch, (**C**) aerial photograph of the entire field (including tomato and other experiments), and (**D**) protective net installed over wheat field (on February 20) to prevent bird influence.

Field trials: tomato yield under reduced irrigation

We also tested the efficacy of SHS mulch under reduced irrigation, where we applied only 50% of the amount of freshwater (Fig. S11). In this case we irrigated the fields only in the morning. Additionally, we tested the application of the SHS as a barrier underground to prevent percolation, adding 20 mm SHS at a depth of about 40–50 cm. Although the results under low irrigation conditions did not show statistical significance ($p > 0.05$) under the Kruskal-Wallis test, we expect to see statistical significance in increasing tomato fruit yields (25%–49%) for a higher number of replicates. The lower effectiveness of the SHS in low irrigation might be linked to the lower saturation of the topsoil in this case as compared to the normal irrigation scenario. The differences between SHS and bare soil were more pronounced only during the morning, after the field had been irrigated, with decreasing effectiveness as the moisture decreased.

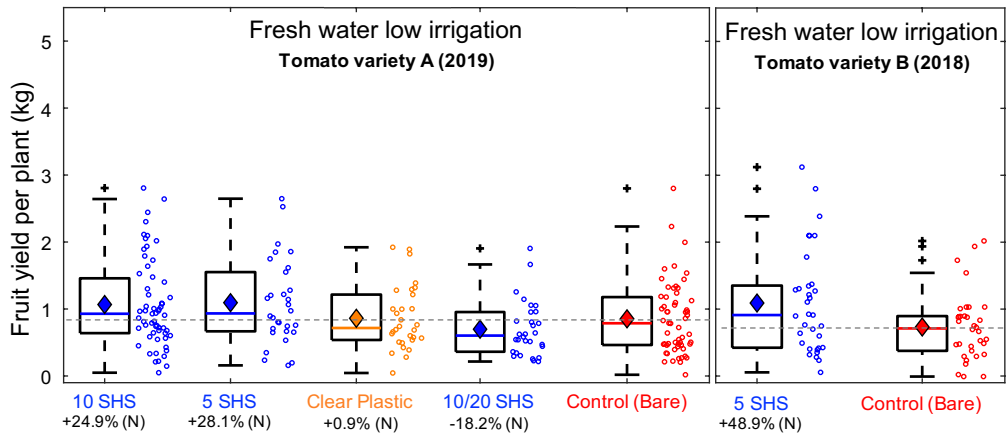

**Fig. S11 – Effects of superhydrophobic sand mulches on crop performance of tomatoes under 50% reduced irrigation.** We reduced the irrigation level by 50% by performing irrigation only once per day during the morning (unlike the normal case, which was twice per day). Code 10/20 SHS represents 10 mm of SHS as a mulch (on top) and 20 mm SHS as a barrier to prevent percolation at about 40–50 cm depth (see Fig. S9E), resulting in higher stress to the plants due to the depth of the lower SHS layer. We did not test black plastic mulch.

Field trials: plant dry biomass

At the end of the crop cycles, we harvested the plants, put them in bags and left them to dry in a greenhouse for three weeks. Next, we measured the individual dry biomass for tomato plants (Fig. S12), and wheat and barley plants (Fig. S13). In general, we found the changes in the plant dry biomass due to the SHS mulches to be less significant than the changes in the plant yields. For tomatoes, we found differences for the tomato variety B (Nunhem's Tristar F1) under fresh (+45%) and brackish (+59%) water irrigation (Fig. S12A). In the low irrigation case, the plant dry biomass for the clear plastic was significantly reduced (-41%) for tomato variety A (Bushra), and significantly increased (47%) for tomato variety B (Nunhem's Tristar F1) (Fig. S12B). There was no significant difference in the dry biomass of wheat, while barley showed a significant increase of 44% under brackish irrigation (Fig. S13).

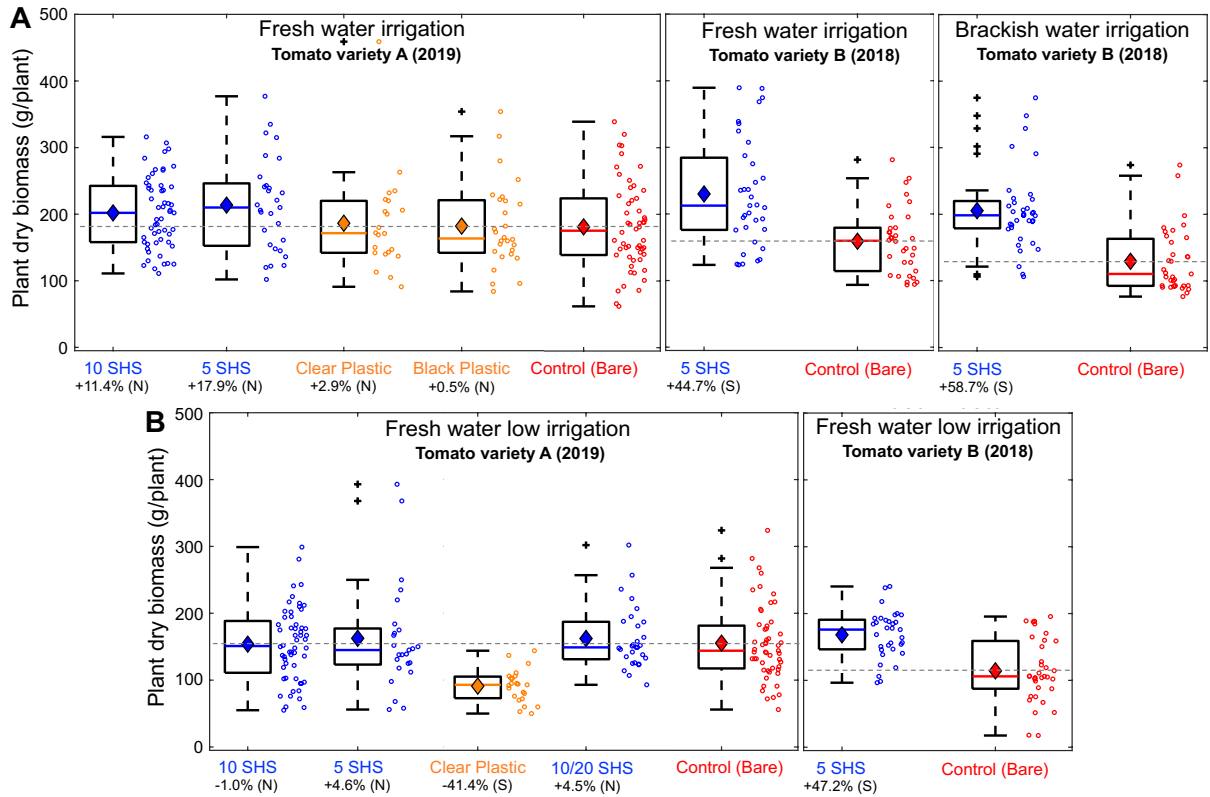

**Fig. S12 – Field results for tomato plant dry biomass.** (**A**) Normal irrigation level and (**B**) 50% reduced irrigation. Dots in the boxplot represent the measurements for individual plants from replicate plots. The number of plots and replicates (n) varied according to the variable. The boxes contain the middle 50% of the data points; the horizontal line indicates the median and the diamond inside the box indicates the mean. We compared each treatment relative to the control (bare) case using the Kruskal-Wallis H test, where (S) represents statistical significance ($p < 0.05$) and the percent change is the relative difference between the means. Tomato (*Solanum lycopersicum*) variety A is Bushra, and variety B is Nunhem's Tristar F1.

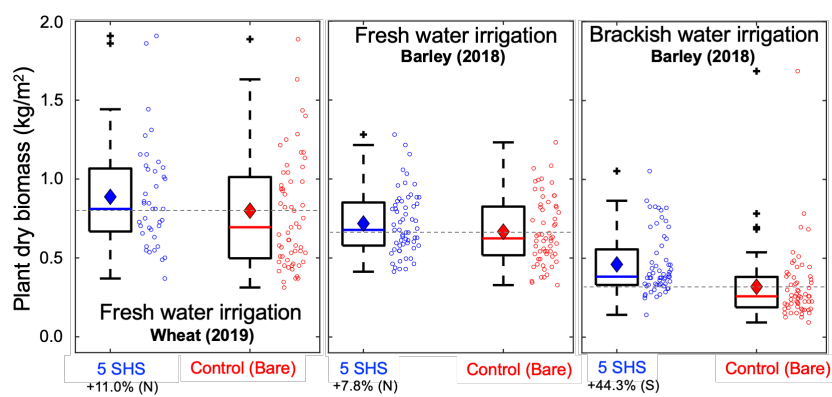

**Fig. S13 – Field results for barley and wheat plant dry mass.** Dots in the boxplot represent the measurements for individual plants from replicate plots. The number of plots and replicates (n) varied according to the variable. The boxes contain the middle 50% of the data points; the horizontal line indicates the median and the diamond inside the box indicates the mean. We compared each treatment relative to the control (bare) case using the Kruskal-Wallis H test, where (S) represents statistical significance ($p < 0.05$) and the percent change is the relative

difference between the means. Wheat (*Triticum aestivum*) variety is Balady, and barley (*Hordeum vulgare*) variety is Morex.

Field trials: tomato yields for intercalated SHS and bare plants

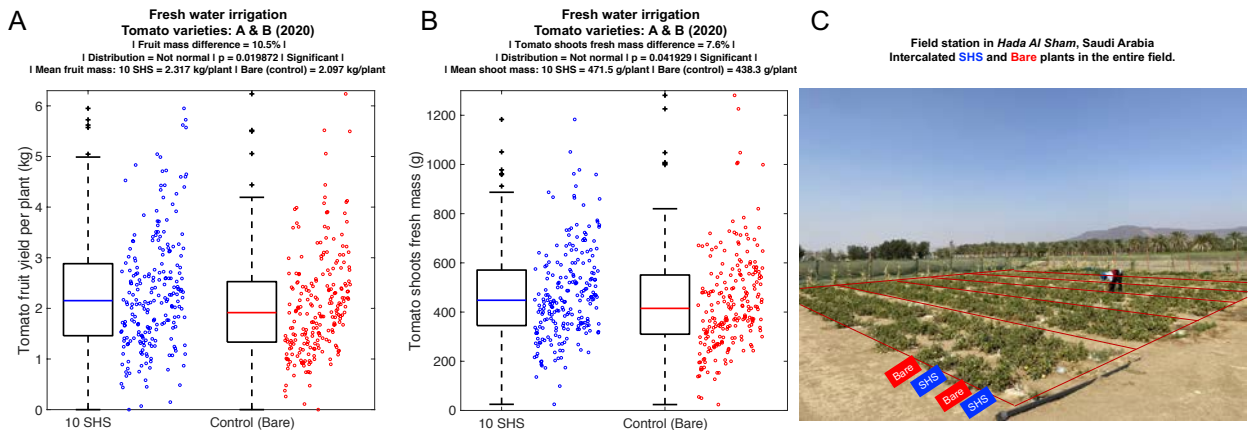

**Fig. S14 – Effects of superhydrophobic sand mulches on crop performance of tomatoes with intercalated SHS and bare plants in 2020.** (**A**) Tomato fruit yield; (**B**) tomato shoots fresh mass. Results for plants under 10 mm superhydrophobic sand (10 SHS) mulches (blue) versus bare soil (red). (**C**) Photograph of the field showing the intercalation of SHS and bare soil plants. The combined moisture retention effect of the SHS next to the bare soil plants decreased the absolute difference between treatment and control. However, the results were still significantly better for SHS with *p* = 0.02. Dots in the boxplot represent the measurements for individual plants. The boxes contain the middle 50% of the data points; the horizontal line indicates the median value; we compared each treatment relative to the control (bare soil) case using the Kruskal-Wallis H test. Differences in both fruit yield and shoots fresh mass were statistically significant ($p < 0.05$). Tomato (*Solanum lycopersicum*) variety A is Bushra, and variety B is Nunhem's Tristar F1; both varieties were intercalated between plots.

# SECTION III – FIELD SOIL CHARACTERIZATION

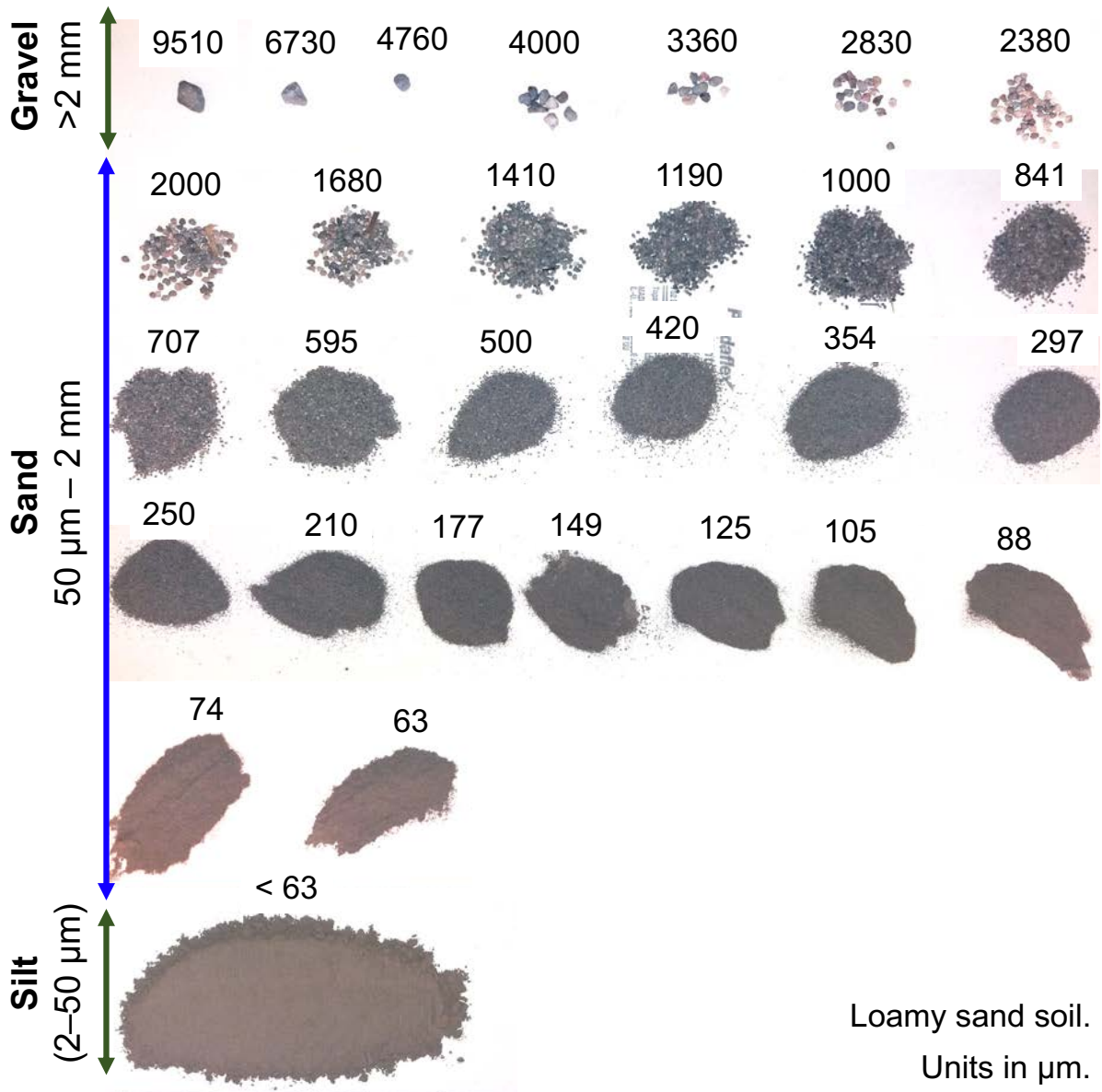

**Fig. S15 – Photograph of soil fractions from sieve analysis for local loamy sand soil from the field station,** *Hada Al Sham*, Saudi Arabia (21.7963° N, 39.7265° E).

Procedure for soil chemical analysis for 2018 season

We collected three soil samples from each plot at different locations within the plot—in total, nine samples per treatment. We removed the top 1 cm layer of the soil, and then collected the samples from 2–10 cm depth, homogeneized them and put them in sealed plastic bags. We identified the samples and shipped them to Geomar Lab (Germany) for analysis.

Geomar performed the analyses according to the protocols mentioned below. In Table S2 we present the results.

*Soil leaches*: We leached the soils with deionized water according to EPA protocols. The procedure involved the leaching of ca. 4 g sediment sample with 45 ml deionized water for a period of 4 hours, followed by centrifugation and subsequent filtration using a 0.45 µm hydrophilic syringe filter (PES, 33 mm)[4].

*Nutrient analyses*: We analyzed nitrate, silicate and phosphate in the deionized water leaches of the soils using standard auto-analyzer techniques on a Seal Quaatro instrument, following the best practice guide for performing nutrient measurements. We analyzed ammonium with a fluorescence method, using Orthophthalate[5].

*pH measurements*: We measured pH in the soil leach solutions using a Metrohm pH meter with an Orion-Thermo Fischer electrode, calibrated using NIST pH buffers.

*Conductivity measurements*: We measured conductivity (in dS/m) in the soil leach solutions using a Metrohm conductivity meter with an Orion-Thermo Fischer electrode, calibrated using Cl solutions. We calculated soil salinity (in ppt or g/L) from the conductivity measurements, which we conducted using a standard approach utilizing 1 g soil with 1 ml added water.

*ICP-OES analysis of soil leachates*: We obtained total elemental concentrations of Ca, Mg, K in the leachate samples using an inductively coupled plasma optical emission spectrometer (ICP-OES; SPECTRO Ciros). We prepared standard solutions from certified single element standard solutions. We corrected all sample values for blank interference by subtracting the mean procedural blank values.

*Particulate Organic Carbon (POC) and Particulate Organic Nitrogen (PON) analysis*: We analyzed particulate carbon and nitrogen samples with an elemental analyzer (Thermo Finnegan Flash EA1112) using acetanilide as the calibration standard. The analysis yielded total particulate carbon and nitrogen. On a separate sample aliquot, we removed inorganic carbon by acidification with sulfurous acid ($H_2SO_3$) under vacuum for 24–48 h, and the subsequent analysis provided POC and PON data[6]. Subtraction of the POC and PON from the total carbon and nitrogen results yielded the particulate inorganic carbon and nitrogen, PIC and PIN respectively. We chose the factor 2.0 to convert POC to POM[7].

*Cation Exchange capacity (CEC)*: We used the hexaamminecobalt (III) chloride method to determine CEC, adding $CaCO_3$ as the pH of the samples was basic. We followed the standard method according to ISO 23470:2018. We weighed the sample of air-dried soil and transferred it to a tightly stoppered polyethylene centrifuge tube of about 50 mL capacity. Then we added 50 mL of 0.0166 mol/L hexaamminecobalt (III) chloride solution to the soil and shook it on the rotary shaker for 1 h. After this we balanced the tubes and centrifuged them at 3000 g for 10 min. We determined the cobalt concentration in the extract by ICP-AES. We calculated the CEC on the basis of the difference between the initial concentration of cobalt in extraction solution and the quantity remaining in extract.

| Season 2018 | Soil analysis tomato and barley fields | pH | Soil salinity | Electric conductivity | Total nitrogen | Total carbon | Organic nitrogen | Organic carbon | Inorganic carbon | % Inorganic carbon | TC/TN ratio | Organic matter | Water holding capacity | CEC | NO$_3$ | NO$_2$ | Silicate | PO$_4$ | Ca | K | Mg | Na |
|---|---|---|---|---|---|---|---|---|---|---|---|---|---|---|---|---|---|---|---|---|---|---|
| | | pH | mg/L | dS/m | mg/kg | mg/kg | mg/kg | mg/kg | mg/kg | % | TC/TN | % | % | mEq/100g | mg/kg | mg/kg | mg/kg | mg/kg | mg/kg | mg/kg | mg/kg | mg/kg |
| TOMATO (fresh) | SHS Mean (n=9) | 7.7 | 11.8 | 22.7 | 1227.9 | 9954.8 | 1110.1 | 8253.2 | 1701.6 | 17.6% | 8.5 | 1.7% | 33.0% | 12.2 | 273.7 | 1.3 | 49.4 | 13.0 | 3763.2 | 511.8 | 419.4 | 983.7 |
| | SHS Std. Error (%) | 0.3% | 7.4% | 7.4% | 11.1% | 7.2% | 10.5% | 8.9% | 12.5% | 14.7% | 7.0% | 8.9% | 2.9% | 2% | 19.9% | 18.8% | 4.6% | 5.8% | 3.7% | 15.1% | 9.7% | 20.8% |
| | Bare Mean (n=9) | 7.7 | 16.2 | 31.2 | 1530.0 | 10991.2 | 949.2 | 7027.5 | 3963.6 | 33.3% | 7.2 | 1.4% | 27.7% | 11.7 | 608.0 | 0.9 | 39.8 | 8.9 | 4515.7 | 354.3 | 680.7 | 2475.3 |
| | Bare Std. Error (%) | 0.4% | 11.2% | 11.2% | 7.4% | 8.3% | 6.6% | 4.6% | 22.2% | 15.7% | 7.0% | 4.6% | 4.2% | 2% | 2.7% | 18.7% | 6.3% | 13.0% | 4.3% | 4.9% | 3.1% | 5.0% |
| | Difference (SHS vs Bare %) | -0.3% | -27.3% | -27.3% | -19.7% | -9.4% | 17.0% | 17.4% | -57.1% | -47.1% | 17.7% | 17.4% | 19.0% | 4.2% | -55.0% | 36.4% | 24.4% | 47.0% | -16.7% | 44.4% | -38.4% | -60.3% |
| | Kruskal-Wallis test (p<0.05) | N | S | S | N | N | N | N | S | S | S | N | N | N | S | N | N | N | S | N | S | S |
| TOMATO (brackish) | SHS Mean (n=9) | 7.6 | 10.9 | 20.9 | 1253.1 | 12921.7 | 1051.7 | 8436.9 | 4484.8 | 32.4% | 10.5 | 1.7% | 37.1% | 12.7 | 164.7 | 1.6 | 54.2 | 18.2 | 3528.0 | 292.3 | 346.7 | 1184.7 |
| | SHS Std. Error (%) | 0.6% | 5.4% | 5.4% | 6.3% | 10.5% | 4.6% | 5.4% | 25.4% | 12.4% | 12.1% | 5.4% | 3.8% | 3% | 8.0% | 24.7% | 2.9% | 16.2% | 6.4% | 7.8% | 5.7% | 8.3% |
| | Bare Mean (n=9) | 7.4 | 15.9 | 30.6 | 1376.5 | 10713.5 | 768.5 | 6011.3 | 4702.2 | 43.0% | 7.9 | 1.2% | 35.5% | 11.9 | 437.7 | 7.2 | 50.9 | 15.3 | 5240.9 | 339.6 | 657.6 | 1945.2 |
| | Bare Std. Error (%) | 0.6% | 5.4% | 5.4% | 5.5% | 4.6% | 3.6% | 4.5% | 12.2% | 8.3% | 5.7% | 4.5% | 3.6% | 2% | 15.3% | 48.2% | 5.6% | 10.0% | 3.3% | 9.3% | 8.5% | 10.8% |
| | Difference (SHS vs Bare %) | 1.6% | -31.6% | -31.6% | -9.0% | 20.6% | 36.9% | 40.4% | -4.6% | -24.7% | 33.5% | 40.4% | 4.6% | 6.7% | -62.4% | -77.8% | 6.6% | 18.8% | -32.7% | -13.9% | -47.3% | -39.1% |
| | Kruskal-Wallis test (p<0.05) | N | S | S | N | N | N | N | N | S | S | S | N | N | S | N | N | N | S | N | S | S |
| BARLEY (fresh) | SHS Mean (n=9) | 7.7 | 10.5 | 20.3 | 861.5 | 8854.6 | 699.3 | 5663.8 | 3190.8 | 34.8% | 10.5 | 1.1% | 32.1% | 11.5 | 95.8 | 0.5 | 56.5 | 16.6 | 4547.0 | 165.3 | 387.9 | 586.6 |
| | SHS Std. Error (%) | 0.3% | 3.7% | 3.7% | 12.4% | 10.4% | 9.5% | 10.9% | 18.0% | 13.1% | 4.3% | 10.9% | 2.1% | 3% | 8.9% | 13.6% | 3.5% | 16.2% | 2.8% | 6.1% | 4.0% | 12.4% |
| | Bare Mean (n=9) | 7.6 | 10.4 | 20.0 | 888.8 | 8769.7 | 652.4 | 4927.8 | 3841.9 | 43.7% | 9.9 | 1.0% | 31.3% | 11.4 | 145.1 | 0.4 | 52.9 | 11.6 | 4485.6 | 175.2 | 344.7 | 434.8 |
| | Bare Std. Error (%) | 0.2% | 4.1% | 4.1% | 5.6% | 4.7% | 6.9% | 7.4% | 8.8% | 7.0% | 3.1% | 7.4% | 4.2% | 2% | 15.0% | 10.3% | 2.3% | 11.0% | 4.3% | 11.2% | 4.0% | 15.9% |
| | Difference (SHS vs Bare %) | 0.8% | 1.2% | 1.2% | -3.1% | 1.0% | 7.2% | 14.9% | -16.9% | -20.5% | 5.8% | 14.9% | 2.4% | 0.7% | -33.9% | 4.6% | 6.8% | 43.7% | 1.4% | -5.7% | 12.5% | 34.9% |
| | Kruskal-Wallis test (p<0.05) | S | N | N | N | N | N | N | N | N | N | N | N | N | N | N | N | N | N | N | N | N |
| BARLEY (brackish) | SHS Mean (n=9) | 7.7 | 14.9 | 28.7 | 1541.6 | 11246.7 | 1077.0 | 7022.7 | 4224.0 | 37.0% | 7.5 | 1.4% | 37.5% | 12.6 | 425.2 | 1.1 | 48.1 | 15.2 | 4660.7 | 331.1 | 602.7 | 1742.5 |
| | SHS Std. Error (%) | 0.6% | 7.1% | 7.1% | 7.9% | 4.0% | 7.8% | 6.4% | 14.1% | 12.5% | 4.7% | 6.4% | 1.6% | 1% | 13.7% | 13.6% | 4.5% | 12.5% | 5.3% | 6.5% | 9.9% | 8.9% |
| | Bare Mean (n=9) | 7.6 | 14.4 | 27.6 | 1855.4 | 15545.3 | 1276.8 | 10341.9 | 5203.4 | 36.6% | 9.0 | 2.1% | 38.8% | 13.1 | 311.9 | 2.5 | 49.7 | 17.9 | 4594.4 | 421.6 | 558.0 | 1715.6 |
| | Bare Std. Error (%) | 0.4% | 9.4% | 9.4% | 24.0% | 19.5% | 21.7% | 25.1% | 14.1% | 11.8% | 4.8% | 25.1% | 3.2% | 2% | 28.5% | 34.1% | 4.4% | 11.7% | 5.1% | 22.0% | 16.3% | 13.5% |
| | Difference (SHS vs Bare %) | 1.2% | 3.9% | 3.9% | -16.9% | -27.7% | -15.6% | -32.1% | -18.8% | 1.1% | -17.2% | -32.1% | -3.2% | -4.0% | 36.3% | -57.7% | -3.2% | ##### | 1.4% | -21.4% | 8.0% | 1.6% |
| | Kruskal-Wallis test (p<0.05) | N | N | N | N | N | N | N | N | N | N | N | N | S | N | N | N | N | N | N | N | N |
| Uncultivated soil | Uncultivated soil (n=3) | 7.7 | 8.8 | 16.9 | 814.9 | 9476.6 | 576.0 | 4424.0 | 5052.6 | 27.0% | 11.7 | 0.9% | 27.7% | 12.8 | 116.8 | 0.4 | 42.0 | 3.2 | 2615.9 | 304.4 | 262.5 | 700.6 |
| | Std. Error (%) | 0.0% | 2.6% | 2.6% | 8.7% | 2.8% | 9.0% | 9.4% | 8.1% | 32.2% | 6.9% | 9.4% | 2.6% | 2% | 2.5% | 4.8% | 1.3% | 5.4% | 3.4% | 3.1% | 2.6% | 2.0% |

**Table S2 – Soil chemical analysis field trial 2018.** (S) indicates statistically significant difference ($p < 0.05$) and (N) indicates no difference ($p < 0.05$) under Kruskal-Wallis test . *Hada Al Sham*, Saudi Arabia (21.7963° N, 39.7265° E).

Procedure for soil chemical analysis for 2019 season

We collected soil samples from different points within each plot and then mixed them into a single bag. We collected replicates from other plots of similar treatments. We removed the top 1 cm layer of soil, and then collected the samples from 2–10 cm depth, homogenized them, and put them in sealed plastic tubes. We labelled the samples and stored them for further analysis at KAUST analytical core laboratory. We conducted the soil analysis during the mid-stage of the crop cycle in March and at the end of the crop cycle in May, 2019. In Tables S3–S4 we present the results.

*Soil leaches*: We leached the soils with a solution of 0.5 M ammonium acetate and 0.5 M acetic acid. For the procedure we used 7 g of sediment sample with 45 ml of solution. We then transferred the samples to a shaker for 30 min and decanted the supernatant. Finally, we filtered the samples with No. 2 filter paper.

*ICP-OES analysis of soil leachates*: We diluted the leached samples 10 times for this procedure, and prepared the standard solution using element standards. We measured the concentrations of Ca, K, Mg, P, S, and Si in the diluted leached samples with an inductively coupled plasma optical emission spectrometer (ICP-OES; PerkinElmer Optima 8300).

*POC and PON analysis*: To determine the total amount nitrogen and carbon in the samples, we used an Elemental Analyzer (Thermo Fisher Scientific Flash 2000) with Soil Standard (Thermo Fisher Scientific – Certificate: 13317) as the calibration standard. First, we measured the weight of each sample to be 10 mg ± 1 mg, placed them in a tin container, sealed them, and prepared them for elemental analysis. Next, we weighted the samples for organic nitrogen (PON) and carbon (POC) by following the procedure explained above, but in this case placing the samples inside of the silver containers. We removed the inorganic part from the samples by adding 5 µL of 3M HCl and kept the samples inside the oven under a temperature of 65 ˚C for 20 min. We repeated this procedure 4 times, in all adding 20 µL of acid solution. thereafter, we dried the samples at 65 ˚C overnight for complete drying. We then sealed the silver containers, wrapped them with tin containers and analyzed them.

| Treatment | | Replicates (n) | | Total nitrogen mg/kg | Total carbon mg/kg | Organic nitrogen mg/kg | Organic carbon mg/kg | Inorganic carbon mg/kg | TC/TN ratio TC/TN | Ca mg/kg | K mg/kg | Mg mg/kg | P mg/kg | S mg/kg | Si mg/kg |
|---|---|---|---|---|---|---|---|---|---|---|---|---|---|---|---|
| TOMATO (Variety A) (March, 2019) (Fresh) | Bare (Control) | 4 | Mean | 1486.0 | 14218.5 | 1021.9 | 8096.6 | 6121.9 | 9.9 | 9867.9 | 1626.4 | 615.5 | 189.6 | 1065.5 | 48.2 |
| | | | Std. Error (%) | 23.2% | 17.3% | 11.5% | 11.5% | 51.2% | 7.9% | 10.5% | 22.7% | 20.6% | 110.3% | 30.7% | 0.183249 |
| | 10 SHS | 4 | Mean | 1119.3 | 11413.9 | 906.5 | 9074.5 | 2339.4 | 10.9 | 10076.8 | 1873.9 | 430.7 | 28.3 | 1382.1 | 49.8 |
| | | | Std. Error (%) | 20.0% | 8.9% | 26.6% | 33.3% | 103.1% | 16.6% | 17.8% | 11.8% | 11.9% | 71.5% | 53.8% | 12.7% |
| | | | Difference vs Bare (%) | -24.7% | -19.7% | -11.3% | 12.1% | -61.8% | 10.6% | 2.1% | 15.2% | -30.0% | -85.1% | 29.7% | 3.3% |
| | | | Kruskal-Wallis test vs bare (p<0.05) | N | N | N | N | N | N | N | N | N | N | N | N |
| | 5 SHS | 2 | Mean | 1028.6 | 11354.2 | 794.2 | 7174.4 | 4179.8 | 11.0 | 9128.6 | 1883.6 | 617.1 | 186.4 | 1025.4 | 45.0 |
| | | | Std. Error (%) | 3.3% | 3.0% | 10.3% | 20.3% | 26.6% | 6.4% | 21.9% | 46.8% | 53.0% | 63.4% | 141.4% | 20.2% |
| | | | Difference vs Bare (%) | -30.8% | -20.1% | -22.3% | -11.4% | -31.7% | 11.7% | -7.5% | 15.8% | 0.3% | -1.7% | -3.8% | -6.7% |
| | | | Kruskal-Wallis test vs bare (p<0.05) | N | N | N | N | N | N | N | N | N | N | N | N |
| | Clear plastic | 2 | Mean | 1131.9 | 12188.7 | 891.9 | 8100.1 | 4088.6 | 10.8 | 7328.6 | 1645.7 | 334.9 | 101.9 | 237.2 | 42.8 |
| | | | Std. Error (%) | 5.1% | 1.1% | 34.4% | 29.2% | 54.6% | 4.0% | 17.4% | 11.6% | 0.3% | 34.4% | 100.0% | 7.4% |
| | | | Difference vs Bare (%) | -23.8% | -14.3% | -12.7% | 0.0% | -33.2% | 9.0% | -25.7% | 1.2% | -45.6% | -46.3% | -77.7% | -11.3% |
| | | | Kruskal-Wallis test vs bare (p<0.05) | N | N | N | N | N | N | N | N | N | N | N | N |
| | Black plastic | 2 | Mean | 853.7 | 12843.2 | 791.5 | 6755.6 | 6087.5 | 15.0 | 11539.3 | 720.0 | 630.0 | – | 2362.5 | 48.2 |
| | | | Std. Error (%) | 5.1% | 8.6% | 17.0% | 10.7% | 6.2% | 3.5% | 22.5% | 24.0% | 18.8% | – | 17.1% | 9.4% |
| | | | Difference vs Bare (%) | -42.5% | -9.7% | -22.5% | -16.6% | -0.6% | 51.9% | 16.9% | -55.7% | 2.3% | | 121.7% | 0.0% |
| | | | Kruskal-Wallis test vs bare (p<0.05) | N | N | N | N | N | N | N | N | N | | N | N |
| WHEAT (March, 2019) (Fresh) | Bare (Control) | 3 | Mean | 872.2 | 10161.4 | 667.0 | 6938.3 | 3223.1 | 11.6 | 11462.1 | 1276.3 | 839.4 | 147.6 | 1801.7 | 38.6 |
| | | | Std. Error (%) | 1.0% | 6.0% | 9.5% | 19.7% | 41.7% | 5.0% | 15.0% | 9.4% | 45.3% | 33.7% | 42.0% | 11.8% |
| | 5 SHS | 2 | Mean | 869.6 | 12868.9 | 777.0 | 6994.0 | 5874.9 | 14.6 | 15383.6 | 1308.2 | 810.3 | 115.7 | 2880.0 | 48.2 |
| | | | Std. Error (%) | 19.2% | 34.8% | 26.1% | 23.7% | 48.0% | 16.1% | 3.7% | 4.5% | 38.9% | 141.4% | 16.4% | 9.4% |
| | | | Difference vs Bare (%) | -0.3% | 26.6% | 16.5% | 0.8% | 82.3% | 25.2% | 34.2% | 2.5% | -3.5% | -21.6% | 59.8% | 25.0% |
| | | | Kruskal-Wallis test vs bare (p<0.05) | N | N | N | N | N | N | N | N | N | N | N | N |
| | Bare (Control) | 4 | Mean | 1992.4 | 13936.9 | 1688.5 | 9331.7 | 4605.2 | 8.3 | 14432.1 | 2277.3 | 769.8 | 321.4 | 1997.7 | 131.8 |
| | | | Std. Error (%) | 32.7% | 13.9% | 30.3% | 8.7% | 28.7% | 24.0% | 4.2% | 5.3% | 10.2% | 37.2% | 46.7% | 7.1% |
| | 10 SHS | 4 | Mean | 1346.6 | 14765.8 | 1308.4 | 9885.3 | 4880.5 | 11.3 | 14416.1 | 1671.4 | 768.2 | 127.0 | 2328.8 | 123.8 |
| | | | Std. Error (%) | 14.8% | 9.4% | 8.5% | 5.2% | 24.9% | 11.0% | 6.9% | 17.9% | 15.5% | 62.3% | 31.5% | 4.5% |
| | | | Difference vs Bare (%) | -32.4% | 5.9% | -22.5% | 5.9% | 6.0% | 36.0% | -0.1% | -26.6% | -0.2% | -60.5% | 16.6% | -6.1% |
| | | | Kruskal-Wallis test vs bare (p<0.05) | N | N | N | N | N | N | N | N | N | N | N | N |
| TOMATO (Variety A) (March, 2019) (Fresh) (Low irrigation) | 5 SHS | 2 | Mean | 1656.0 | 14458.9 | 1311.0 | 9631.4 | 4827.5 | 8.8 | 15750.0 | 1530.0 | 855.0 | 170.4 | 2568.2 | 128.6 |
| | | | Std. Error (%) | 2.5% | 19.9% | 30.9% | 16.2% | 91.9% | 22.4% | 4.6% | 48.7% | 35.1% | 24.0% | 17.5% | 14.1% |
| | | | Difference vs Bare (%) | -16.9% | 3.7% | -22.4% | 3.2% | 4.8% | 5.4% | 9.1% | -32.8% | 11.1% | -47.0% | 28.6% | -2.4% |
| | | | Kruskal-Wallis test vs bare (p<0.05) | N | N | N | N | N | N | N | N | N | N | N | N |
| | Clear plastic | 2 | Mean | 1257.6 | 12744.4 | 1374.2 | 10704.8 | 2040.0 | 10.1 | 16328.6 | 1382.1 | 1542.9 | 305.4 | 3053.6 | 122.1 |
| | | | Std. Error (%) | 1.8% | 0.2% | 13.8% | 1.1% | 6.8% | 2.0% | 8.9% | 29.6% | 23.6% | 93.8% | 4.5% | 7.4% |
| | | | Difference vs Bare (%) | -36.9% | -8.6% | -18.6% | 14.7% | -55.7% | 22.0% | 13.1% | -39.3% | 100.4% | -5.0% | 52.9% | -7.3% |
| | | | Kruskal-Wallis test vs bare (p<0.05) | N | N | N | N | N | N | N | N | N | N | N | N |
| | 10/20 SHS | 2 | Mean | 700.1 | 11750.6 | 939.7 | 7944.8 | 3805.8 | 15.8 | 15557.1 | 1301.8 | 755.4 | 96.4 | 2925.0 | 125.4 |
| | | | Std. Error (%) | 42.2% | 67.5% | 4.4% | 3.0% | 214.5% | 29.4% | 1.8% | 75.1% | 35.5% | 94.3% | 14.0% | 18.1% |
| | | | Difference vs Bare (%) | -64.9% | -15.7% | -44.3% | -14.9% | -17.4% | 90.2% | 7.8% | -42.8% | -1.9% | -70.0% | 46.4% | -4.9% |
| | | | Kruskal-Wallis test vs bare (p<0.05) | N | N | N | N | N | N | N | N | N | N | N | N |

**Table S3 – Soil chemical analysis field trial 2019 (March).** (S) indicates statistically significant difference ($p < 0.05$) and (N) indicates no difference ($p < 0.05$) under Kruskal-Wallis test. *Hada Al Sham*, Saudi Arabia (21.7963° N, 39.7265° E).

| Treatment | Soil analysis tomato and barley fields | Replicates (n) | Total nitrogen mg/kg | Total carbon mg/kg | Organic nitrogen mg/kg | Organic carbon mg/kg | Inorganic carbon mg/kg | TC/TN ratio | Ca mg/kg | K mg/kg | Mg mg/kg | P mg/kg | S mg/kg | Si mg/kg |
|---|---|---|---|---|---|---|---|---|---|---|---|---|---|---|
| **TOMATO (Variety A) (May, 2019) (Fresh)** | Bare (Control) Mean | 4 | 1299.3 | 10464.3 | 1023.2 | 8642.0 | 1822.3 | 9.9 | 12664.3 | 961.1 | 607.5 | 718.4 | 2298.2 | 65.9 |
| | Std. Error (%) | | 35.3% | 6.6% | 10.9% | 6.8% | 5.6% | 24.9% | 4.2% | 21.5% | 13.3% | 34.7% | 24.2% | 7.1% |
| | 10 SHS Mean | 4 | 926.9 | 11727.5 | 884.4 | 7999.6 | 3727.9 | 12.7 | 13178.6 | 739.3 | 613.9 | 376.1 | 2280.5 | 46.6 |
| | Std. Error (%) | | 12.4% | 10.6% | 4.6% | 4.6% | 29.9% | 4.0% | 5.1% | 9.6% | 6.8% | 24.9% | 21.3% | 35.8% |
| | Difference vs Bare (%) | | -29% | 12% | -14% | -7% | 105% | 29.2% | 4.1% | -23.1% | 1.1% | -47.7% | -0.8% | -29.3% |
| | Kruskal-Wallis test vs bare ($p<0.05$) | | N | N | N | N | S | N | N | N | N | N | N | N |
| | 5 SHS Mean | 2 | 1563.2 | 10109.6 | 1175.8 | 8041.2 | 2068.4 | 8.1 | 14560.7 | 932.1 | 649.3 | 675.0 | 2732.1 | 64.3 |
| | Std. Error (%) | | 64.7% | 2.2% | 53.9% | 15.7% | 50.1% | 63.0% | 9.1% | 34.1% | 26.6% | 41.8% | 11.6% | 42.4% |
| | Difference vs Bare (%) | | 20.3% | -3.4% | 14.9% | -7.0% | 13.5% | -17.6% | 15.0% | -3.0% | 6.9% | -6.0% | 18.9% | -2.4% |
| | Kruskal-Wallis test vs bare ($p<0.05$) | | N | N | N | N | N | N | N | N | N | N | N | N |
| | Clear plastic Mean | 2 | 813.0 | 10661.1 | 760.0 | 6819.9 | 3841.2 | 13.1 | 14239.3 | 745.7 | 745.7 | 565.7 | 3375.0 | 73.9 |
| | Std. Error (%) | | 9.2% | 17.6% | 0.5% | 3.7% | 55.4% | 8.5% | 5.4% | 40.2% | 7.3% | 51.4% | 9.4% | 6.1% |
| | Difference vs Bare (%) | | -37.4% | 1.9% | -25.7% | -21.1% | 110.8% | 32.6% | 12.4% | -22.4% | 22.8% | -21.3% | 46.9% | 12.2% |
| | Kruskal-Wallis test vs bare ($p<0.05$) | | N | N | N | N | N | N | N | N | N | N | N | N |
| | Black plastic Mean | 2 | 1030.4 | 13229.0 | 734.4 | 6208.8 | 7020.3 | 13.1 | 13789.6 | 604.3 | 665.4 | 568.9 | 2860.7 | 70.7 |
| | Std. Error (%) | | 43.4% | 34.5% | 6.3% | 7.2% | 58.6% | 9.7% | 6.3% | 24.1% | 13.0% | 50.3% | 14.3% | 0.0% |
| | Difference vs Bare (%) | | -20.7% | 26.4% | -28.2% | -28.2% | 285.2% | 33.1% | 8.9% | -37.1% | 9.5% | -20.8% | 24.5% | 7.3% |
| | Kruskal-Wallis test vs bare ($p<0.05$) | | N | N | N | N | N | N | N | N | N | N | N | N |
| **WHEAT (May, 2019) (Fresh)** | Bare (Control) Mean | 3 | 1051.3 | 8843.8 | 1026.4 | 7527.6 | 1316.2 | 9.5 | 14978.6 | 642.9 | 642.9 | 572.1 | 4178.6 | 70.7 |
| | Std. Error (%) | | 33.3% | 5.6% | 26.0% | 11.5% | 103.4% | 26.1% | 9.6% | 693.3% | 368.0% | 330.8% | 323.5% | 330.3% |
| | 5 SHS Mean | 2 | 878.7 | 9904.6 | 742.1 | 6447.0 | 3457.6 | 12.6 | 13275.0 | 662.1 | 597.9 | 639.6 | 3342.9 | 54.6 |
| | Std. Error (%) | | 49.1% | 8.5% | 41.8% | 22.2% | 17.0% | 41.5% | 17.5% | 23.3% | 25.9% | 29.1% | 24.5% | 8.3% |
| | Difference vs Bare (%) | | -16.4% | 12.0% | -27.7% | -14.4% | 162.7% | 32.5% | -11.4% | 3.0% | -7.0% | 11.8% | -20.0% | -22.7% |
| | Kruskal-Wallis test vs bare ($p<0.05$) | | N | N | N | N | N | N | N | N | N | N | N | N |
| **TOMATO (Variety A) (May, 2019) (Fresh) (Low irrigation)** | Bare (Control) Mean | 4 | 1186.8 | 13109.2 | 1256.8 | 9211.9 | 3897.3 | 11.3 | 12792.9 | 675.0 | 996.4 | 321.4 | 3310.7 | 80.4 |
| | Std. Error (%) | | 18.1% | 13.7% | 32.5% | 44.5% | 67.9% | 7.3% | 2.6% | 7.1% | 19.6% | 17.3% | 2.7% | 4.6% |
| | 10 SHS Mean | 4 | 961.2 | 11942.3 | 878.4 | 6413.0 | 5529.3 | 12.5 | 11282.1 | 458.0 | 628.4 | 157.5 | 2314.3 | 70.7 |
| | Std. Error (%) | | 12.0% | 12.0% | 11.2% | 38.6% | 55.0% | 5.3% | 8.4% | 19.1% | 16.2% | 66.7% | 21.1% | 11.3% |
| | Difference vs Bare (%) | | -19.0% | -8.9% | -30.1% | -30.4% | 41.9% | 10.2% | -11.8% | -32.1% | -36.9% | -51.0% | -30.1% | -12.0% |
| | Kruskal-Wallis test vs bare ($p<0.05$) | | N | N | N | N | N | N | N | N | N | N | N | N |
| | 5 SHS Mean | 2 | 1597.3 | 10390.4 | 737.6 | 3660.0 | 6730.5 | 5.8 | 9771.4 | 739.3 | 543.2 | 395.4 | 1607.2 | 80.4 |
| | Std. Error (%) | | 29.7% | 99.1% | 47.2% | 141.4% | 76.1% | 81.4% | 18.6% | 18.4% | 9.2% | 86.2% | 84.9% | 5.7% |
| | Difference vs Bare (%) | | 34.6% | -20.7% | -41.3% | -60.3% | 72.7% | -48.6% | -23.6% | 9.5% | -45.5% | 23.0% | -51.5% | 0.0% |
| | Kruskal-Wallis test vs bare ($p<0.05$) | | N | N | N | N | N | N | N | N | N | N | N | N |
| | Clear plastic Mean | 2 | 1700.7 | 13544.7 | 1044.4 | 6766.2 | 6778.6 | 8.2 | 12246.4 | 835.7 | 835.7 | 305.4 | 2925.0 | 70.7 |
| | Std. Error (%) | | 36.0% | 19.6% | 15.0% | 29.5% | 68.6% | 17.0% | 1.9% | 0.0% | 43.5% | 34.2% | 1.6% | 12.9% |
| | Difference vs Bare (%) | | 43.3% | 3.3% | -16.9% | -26.5% | 73.9% | -27.3% | -4.3% | 23.8% | -16.1% | -5.0% | -11.7% | -12.0% |
| | Kruskal-Wallis test vs bare ($p<0.05$) | | N | N | N | N | N | N | N | N | N | N | N | N |
| | 10/20 SHS Mean | 2 | 868.6 | 12397.3 | 809.2 | 7439.5 | 4957.9 | 14.4 | 11635.7 | 520.7 | 768.2 | 350.4 | 2539.3 | 70.7 |
| | Std. Error (%) | | 26.5% | 19.9% | 34.7% | 32.3% | 1.5% | 6.7% | 9.4% | 33.2% | 36.1% | 92.1% | 1.8% | 0.0% |
| | Difference vs Bare (%) | | -26.8% | -5.4% | -35.6% | -19.2% | 27.2% | 27.5% | -9.0% | -22.9% | -22.9% | 9.0% | -23.3% | -12.0% |
| | Kruskal-Wallis test vs bare ($p<0.05$) | | N | N | N | N | N | N | N | N | N | N | N | N |

**Table S4 – Soil chemical analysis field trial 2019 (May).** (S) indicates statistically significant difference ($p < 0.05$) and (N) indicates no difference ($p < 0.05$) under Kruskal-Wallis test. *Hada Al Sham*, Saudi Arabia (21.7963° N, 39.7265° E).

Analysis of the bacterial communities associated with the bulk soil and the root systems of barley and tomato

**Table S5. Tomato and barley microbial richness under fresh and brackish water irrigation, and SHS versus bare soil.** We calculated the number of samples, mean (±SD) of sequences, and mean (±SD) of richness (number of SVs) for bacterial communities associated with the root-system compartments (root tissues, rhizosphere and bulk soil) subjected to the different irrigation types (fresh and brackish water) and overlay treatments (bare soil and SHS mulch).

| Crop | Compartment | Irrigation | Overlay | N. sample* | N. seq.* | Richness* |
|---|---|---|---|---|---|---|
| Barley | Bulk soil | Fresh | Bare soil | 9 | 71139 ± 17288 | 1099 ± 131 |
| | | | SHS mulch | 8 | 68014 ± 13005 | 1109 ± 193 |
| | | Brackish | Bare soil | 9 | 63915 ± 9533 | 816 ± 112 |
| | | | SHS mulch | 9 | 54156 ± 20706 | 882 ± 234 |
| | Rhizosphere | Fresh | Bare soil | 8 | 47714 ± 10961 | 557 ± 108 |
| | | | SHS mulch | 9 | 48033 ± 13583 | 660 ± 108 |
| | | Brackish | Bare soil | 7 | 51348 ± 25300 | 665 ± 261 |
| | | | SHS mulch | 9 | 35989 ± 11357 | 535 ± 186 |
| | Root tissues | Fresh | Bare soil | 8 | 163531 ± 51171 | 427 ± 121 |
| | | | SHS mulch | 9 | 141735 ± 55532 | 390 ± 39 |
| | | Brackish | Bare soil | 8 | 141293 ± 56799 | 365 ± 125 |
| | | | SHS mulch | 8 | 144158 ± 63975 | 321 ± 120 |
| Tomato | Bulk soil | Fresh | Bare soil | 9 | 38053 ± 15207 | 478 ± 78 |
| | | | SHS mulch | 8 | 36376 ± 32606 | 421 ± 173 |
| | | Brackish | Bare soil | 7 | 4607 ± 4874 | 79 ± 47 |
| | | | SHS mulch | 9 | 7649 ± 18287 | 81 ± 92 |
| | Rhizosphere | Fresh | Bare soil | 9 | 37442 ± 8514 | 583 ± 208 |
| | | | SHS mulch | 9 | 47854 ± 9694 | 677 ± 175 |
| | | Brackish | Bare soil | 9 | 49944 ± 10258 | 574 ± 161 |
| | | | SHS mulch | 9 | 35890 ± 5878 | 507 ± 107 |
| | Root tissues | Fresh | Bare soil | 8 | 53305 ± 14563 | 862 ± 156 |
| | | | SHS mulch | 8 | 55451 ± 15428 | 981 ± 286 |
| | | Brackish | Bare soil | 8 | 74404 ± 22545 | 1065 ± 249 |
| | | | SHS mulch | 9 | 50498 ± 11575 | 798 ± 147 |

* Values obtained after quality filtering described in the material and methods section and chloroplast/mitochondria removal.

**Table S6 – Results of multiple comparison tests of bacterial compositional similarity of the communities associated with the different root system compartments (root tissues, rhizosphere and bulk soil) in both barley and tomato**. For each fraction, we also evaluated the effect of different irrigation regimes (fresh and brackish water) and overlay treatment (bare soil/superhydrophobic sand mulches). We consider results of t-test with $p < 0.05$ to be significant.

| Crop | Compartment* | Compartment$ | Irrigation# | Irrigation$ | Overlay# |
|---|---|---|---|---|---|
| Barley | $F_{2,98}=18.04$, $p = 0.001$ | Bulk soil (a) | $t = 1.69, p = 0.002$ | Fresh (a) | $t = 1.09, p = 0.257$ |
| | | | | Brackish (b) | $t = 0.89, p = 0.631$ |
| | | Rhizosphere (b) | $t = 2.23, p = 0.001$ | Fresh (a) | $t = 1.18, p = 0.165$ |
| | | | | Brackish (b) | $t = 1.14, p = 0.237$ |
| | | Root tissues (c) | $t = 2.34, p = 0.001$ | Fresh (a) | $t = 1.27, p = 0.079$ |
| | | | | Brackish (b) | $t = 1.14, p = 0.226$ |
| Tomato | $F_{2,98}=14.99$, $p = 0.001$ | Bulk soil (a) | $t = 1.62, p = 0.003$ | Fresh (a) | $t = 1.01, p = 0.451$ |
| | | | | Brackish (b) | $t = 0.91, p = 0.546$ |
| | | Rhizosphere (b) | $t = 2.63, p = 0.001$ | Fresh (a) | $t = 1.34, p = 0.065$ |
| | | | | Brackish (b) | $t = 1.31, p = 0.071$ |
| | | Root tissues (c) | $t = 3.07, p = 0.001$ | Fresh (a) | $t = 1.05, p = 0.342$ |
| | | | | Brackish (b) | $t = 0.86, p = 0.684$ |

\* Results of ANOVA main test
# Results of t-test
$ Lowercase letters in parenthesis indicate the results of multiple pair-wise comparison test among the levels of the experimental factor analyzed

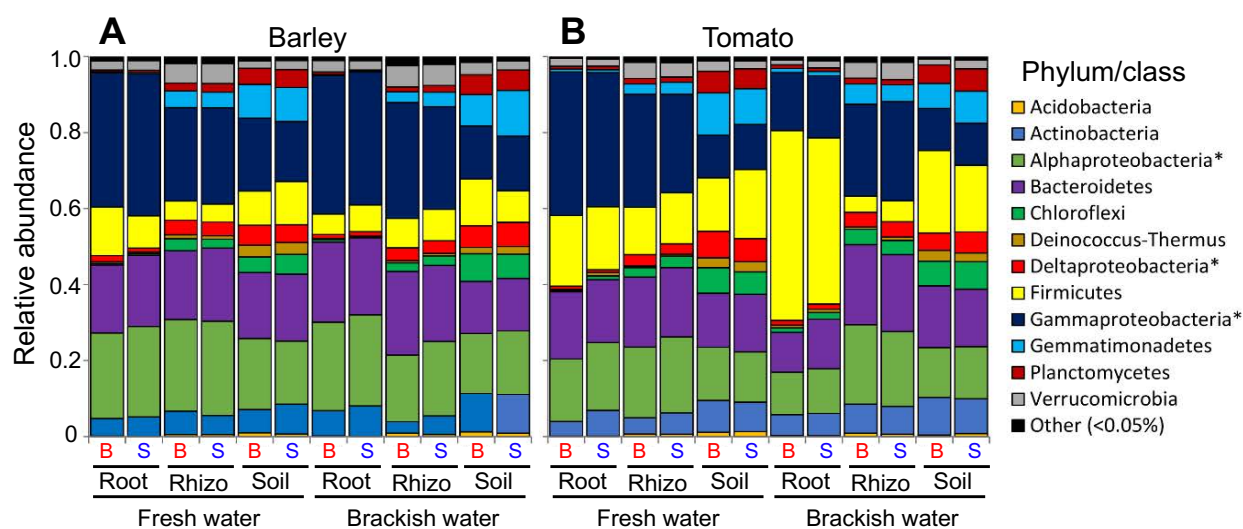

**Fig. S16 – Relative abundance of bacterial phyla/classes for 2018 crop season.** (**A**) Barley and (**B**) tomato. Red B for bare soil and blue S for SHS. (\*) indicates classes belonging to the *Proteobacteria* phylum. Phyla/classes with a relative abundance < 0.01% are classified as others.

**Caption for Movie S1 – Water droplets falling on common sand (left) and superhydrophobic sand (right).** High-speed videos of a water droplets (~2 mm) landing on a packed bed of superhydrophobic sand, released from a height of ~2 cm. Water droplet penetrates into the common sand (left), while another droplet bounces off of the SHS, carrying some SHS particles attached to its interface.

**Caption for Movie S2 – Application of SHS mulches in tomato field.** We used a rubber tube as reference to apply a weigheted amount of SHS over a specific area of 40 cm radius to render the desired mulch thickness.